\documentclass{aa}  

\usepackage{natbib}
\usepackage{graphicx}
\usepackage{txfonts}
\usepackage{hyperref}
\usepackage[dvipsnames]{xcolor}
\usepackage{siunitx} 
\newcommand{\hei}{ \ion{He}{i}\ }
\newcommand{\heii}{ \ion{He}{ii}\ }

\newcommand{\update}[1]{{\color{black}#1}}

\begin{document} 

   \title{X-Shooting ULLYSES: Massive Stars at low metallicity}
  \subtitle{VIII. Stellar and wind parameters of newly revealed stripped stars in Be binaries}

  \author{V. Ramachandran\inst{\ref{inst:ari}} \and A.A.C. Sander\inst{\ref{inst:ari}}  \and D. Pauli\inst{\ref{inst:up}}  \and J. Klencki\inst{\ref{inst:ESO}} \and F. Backs\inst{\ref{inst:api}} \and F. Tramper\inst{\ref{inst:kul}}   
 \and M. Bernini-Peron\inst{\ref{inst:ari}} \and P. Crowther\inst{\ref{inst:shef}}  \and W.-R. Hamann\inst{\ref{inst:up}} \and R. Ignace\inst{\ref{inst:TS}} \and R. Kuiper\inst{\ref{inst:DB}} \and S. Oey\inst{\ref{inst:mich}}  \and L. M. Oskinova\inst{\ref{inst:up}} \and T. Shenar\inst{\ref{inst:isl}} \and H. Todt\inst{\ref{inst:up}} \and J.S. Vink\inst{\ref{inst:aop}}  \and L. Wang\inst{\ref{inst:cas}} \and A. Wofford\inst{\ref{inst:mex}} \and the XShootU collaboration
          }

   \institute{
      {Zentrum f{\"u}r Astronomie der Universit{\"a}t Heidelberg, Astronomisches Rechen-Institut, M{\"o}nchhofstr. 12-14, 69120 Heidelberg\label{inst:ari}}\\
              \email{vramachandran@uni-heidelberg.de} 
     \and         {Institut f{\"u}r Physik und Astronomie, Universit{\"a}t Potsdam, Karl-Liebknecht-Str. 24/25, D-14476 Potsdam, Germany\label{inst:up}} 
    \and {European Southern Observatory, Karl-Schwarzschild-Strasse 2, 85748, Garching bei München, Germany \label{inst:ESO} }
     \and         {Anton Pannekoek Institute for Astronomy, Universiteit van Amsterdam, Science Park 904, 1098 XH Amsterdam, The Netherlands\label{inst:api}}  
     \and         {Institute of Astronomy, KU Leuven, Celestijnenlaan 200D, B-3001 Leuven, Belgium\label{inst:kul}}
     \and {Dept of Physics \& Astronomy, University of Sheffield, Hounsfield Road, Sheffield S3 7RH, UK  \label{inst:shef}}
     \and {Department of Physics \& Astronomy, East Tennessee State University, Johnson City, TN 37614, USA \label{inst:TS}}
     \and {Fakultät für Physik, Universität Duisburg-Essen, Lotharstra{\ss}e 1, 47057 Duisburg      \label{inst:DB}}       
     \and {University of Michigan, Department of Astronomy, 323 West Hall, Ann Arbor, MI 48109, USA \label{inst:mich} }
     \and {The School of Physics and Astronomy, Tel Aviv University, Tel Aviv 6997801, Israel \label{inst:isl} }
     \and        {Armagh Observatory and Planetarium, College Hill, BT61 9DG Armagh, Northern Ireland \label{inst:aop}}
     \and {Yunnan Observatories, Chinese Academy of Sciences, Kunming 650216 Yunnan, PR China \label{inst:cas}}
      \and {Instituto de Astronomía, Universidad Nacional Autónoma de México, Unidad Académica en Ensenada, Km 103 Carr. TijuanaEnsenada, Ensenada, B.C. CP 22860, Mexico \label{inst:mex}}
             }

\date{}

\abstract{
On the route towards merging neutron stars and stripped-envelope supernovae, binary population synthesis predicts a large number of post-interaction
systems with massive stars that have been stripped off their outer layers. Yet, observations of such stars in the intermediate-mass regime below the Wolf-Rayet masses are rare.
Using X-Shooting ULLYSES (XShootU) data, we have discovered three partially stripped star + Be/Oe binaries in the Magellanic Clouds. We analyzed the UV and optical spectra using the Potsdam Wolf-Rayet (PoWR) model atmosphere code by superimposing model spectra that correspond to each component. The estimated current masses of the partially stripped stars fall within the intermediate mass range of $\approx 4-8\,M_{\odot}$. These objects are found to be over-luminous for their corresponding stellar masses, which aligns with the luminosities during core He-burning. Their accompanying Be/Oe secondaries are found to have much higher masses than their stripped primaries (mass ratio $\gtrsim 2$). 
The surfaces of all three partially stripped stars exhibit clear indications of significant nitrogen enrichment as well as a depletion of carbon and oxygen. Furthermore, one of our sample stars shows signs of substantial helium enrichment. Our study provides the first comprehensive determination of the wind parameters of partially stripped stars in the intermediate mass range. The wind mass-loss rates of these stars are estimated to be on the order of $10^{-7}\,M_\odot\, \mathrm{yr}^{-1}$, which is more than ten times higher than that of OB stars with the same luminosity. The current mass-loss recipes commonly employed in evolutionary models to characterize this phase are based on OB or WR mass-loss rates, and they significantly underestimate or overestimate the observed mass-loss rates of (partially) stripped stars by an order of magnitude. 
Binary evolution models suggest that the observed primaries had initial masses in the range of 12--17\,$M_{\odot}$, and are potential candidates for stripped-envelope supernovae resulting in the formation of a neutron star. If these systems survive the explosion, they will likely evolve to become Be X-ray binaries and later double neutron stars.
}

  \keywords{massive star -- binaries: spectroscopic --stars:fundamental parameters}

\maketitle


\section{Introduction}

Massive stars are key players in the evolution of the Universe, as they produce heavy elements, emit ionizing radiation, explode as supernovae, and collapse into compact objects. However, there are still major gaps in our understanding on their evolution, fate, and impact to their surroundings, especially at low metallicity, which hampers our view on the formation of gravitational wave sources and the re-ionization of the universe. 

One of the key factors affecting the evolution and final remnant mass of massive stars is binary interaction. Most massive stars are not isolated but belong to binary or multiple systems, where they can exchange mass and angular momentum with their companions \citep{Sana2012,Sana2014}. Binary evolution can lead to the formation of a significant population of core helium-burning stars that have lost a significant part of their hydrogen-rich envelope after mass transfer via Roche-lobe overflow. These stars are known as stripped-envelope stars, and their spectral characteristics vary from hot subdwarfs to Wolf-Rayet (WR) stars, depending on their masses \citep[e.g.,][]{Paczynski1967,Vanbeveren1991,Eldridge2008,gotberg_2017_ionizing}. However, stripped-envelope stars with masses between the regimes of low-mass subdwarfs ($\lesssim 1.5\,M_\odot$) and classical WR stars ($\gtrsim 8\,M_\odot$) are rarely observed. Furthermore, the minimum mass of WR stars is found to increase with lower metallicities \citep[ $\gtrsim 17\,M_\odot$ at Z=0.2\,$Z_{\odot}$;][]{Shenar2020}  expanding the range of intermediate masses.
\citet{Drout2023Sci} used UV photometry and follow-up optical spectroscopy in the Magellanic Clouds to specifically look for hot and compact phases of stripped stars. \citet{Gotberg2023} carried out a spectroscopic analysis of a sub-sample of these stripped stars, which revealed seven hot He stars in the intermediate mass regime. However, they found that the contribution from the secondary companion (mass gainer) to the optical spectra in their sub-sample is negligible. They thus suggest that these hot stripped He stars are the result of either common envelope ejection in the case secondary is a main sequence star, or -- in the case of compact companions -- stable mass transfer or common envelope ejection. Where as in the lower mass range a handful of subdwarf O + Be binaries have been observed in the Galaxy \cite[e.g.,][]{Schootemeijer2018,Wang2021}.

Binary evolution models predict that these stripped He stars are typically fainter than their main-sequence companion and, hence, can be easily outshined \citep[for e.g.,][]{Yungelson2024}. This makes their identification not easily achieved without the use of detailed quantitative spectroscopy. There have been a few recent detections suggesting that the predicted population of stripped-envelope stars, which has so far been mostly absent in observations, may actually be present within the known population of OB stars. Systems originally proposed to be X-ray quiet black hole + Be binaries such as LB1 and HR6819 \citep{Liu2019,Rivinius2020} have been revealed as partially stripped star + Be binaries \citep[e.g.,][]{Abdul-Masih2020,Bodensteiner2020HR6819,Shenar2020lb1,El-Badry2021, Frost2022}. \citet{Ramachandran2023} recently discovered the first partially stripped massive star in a binary with a massive Be-type companion located in the low metallicity environment of the Small Magellanic Cloud (SMC). \citet{Villasenor2023} has identified a similarly bloated stripped star with an early B-type companion in the Large Magellanic Cloud (LMC). Yet, most of these discovered stripped stars in binaries seem to not reside in the predicted long-lived stage of core He burning on the He-Zero age main sequence, where they are hot and fully stripped. Contrarily, observations find cooler, partially stripped stars, which are associated with short transition phases. 
 
 Recent evolutionary models by \citet{Klencki2022} suggest that binary evolution at low metallicity might favor partial envelope stripping and mass transfer on a long, nuclear time scale. The presence or absence of a compact stripped-star population would have a severe impact on population synthesis predictions, for example, due to the different evolutionary fates and ionizing fluxes. Accurate knowledge about the evolution of massive binary stars is needed to explain observed gravitational wave mergers. However, there is a significant lack of observations for binary systems containing a massive main sequence star with a stripped star companion. Whether this is due to observational challenges or errors in the underlying evolutionary models is one of the key questions in massive star research. A particular decisive, but essentially unknown, ingredient are the wind mass-loss rates of these intermediate-mass partially stripped stars, which can only be constrained with UV spectroscopy. There are currently no studies on the winds of stripped stars due to the reliance on optical spectra in the newly discovered systems, which lack UV spectra. 
 
 In this paper, we report the discovery of three partially stripped stars with Oe/Be companions in the Magellanic Clouds using multi-epoch and multi-wavelength spectroscopy.
 \update{We refer to the mass donor, the stripped star, as "primary" throughout this work, and the mass gainer, the spun-up star, as "secondary".}
 To search for ``hidden'' stripped stars among the known OB population, we used both UV and optical spectra provided by the ``X-Shooting ULLYSES'' (XShootU) project\footnote{https://massivestars.org/xshootu/}. The XShootU data cover hundreds of hot massive stars in the Magellanic Clouds, providing a good sample to identify and characterize stripped stars in binaries among the OB populations at low metallicity. 
 
 As a result of binary interaction (stable mass transfer), the secondary (i.e., the initially less massive) component gains mass and angular momentum \citep{Kriz1975,Pols1991,Langer2020}. Such secondaries would evolve into rapidly rotating stars \citep[e.g.,][]{deMink2013,Renzo2021}, which can have disk emission features and appear as Be stars \citep{Pols1991,Shao2014,Bodensteiner+2020MS}. Moreover, the lack of main sequence companions to Be stars \citep{Bodensteiner+2020MS}, the evidence from SEDs for a tidal truncation of the Be disk by a companion \citep{Klement2019}, as well as population synthesis predictions \citep{Shao2014,Hastings2021}, indicate that Be stars are part of post-interaction binaries, and are the best candidates to look for stripped stars companions. 
 We thus focused on objects within the XShootU sample that show disk emission features (Oe/Be) in their optical spectra. We looked for the spectral characteristics of binaries with stripped stars and Be stars, as suggested by \cite{Ramachandran2023}. 
 We carefully inspect narrow and broad components of stellar absorption lines in such systems and look for radial velocity variations if multi-epoch observations are available. In a sub-sample of these systems, we carried out the spectral analysis and confirmed the detection of three systems consisting of a partially stripped star companion and a Be/Oe star which we present in this work. 
 By analyzing both the UV and optical spectra, we are providing the empirical wind parameters of the stripped stars for the first time. The observations used in this study are described in Sect.\,\ref{observation}. The details on the spectroscopic analysis and parameters derived for each system are covered in Sect.\,\ref{2dFS163}, \ref{2dFS2553} and \ref{SK-7135}. Discussions on mass, chemical abundances, wind, and evolution of the stripped stars are presented in Sect.\,\ref{discussion}, followed by conclusions in Sect.\,\ref{summary}. In Sect.\,\ref{appendix:2dfs2553} to \ref{appendix:plots} details of the orbital analysis, evolution models, and additional plots are given.

\section{Observations}
\label{observation}
The objects are part of the XShootU sample. The details of the project, observations, and data are explained in \citep{Vink2023, Roman-Duval2020}. 
For two of the sample stars (2dFS\,163 and \mbox{Sk\,-71$^{\circ}$\,35}) UV spectra are taken with the Hubble Space Telescope's Cosmic Origins Spectrograph (HST/COS) with the G130M/1291  and G160M/1611  gratings.
For 2dFS\,2553, the UV spectrum is obtained using the Space Telescope Imaging Spectrograph (HST/STIS) with the E140M/1425.
Additionally, we utilized the available HST/COS spectra taken with the G130M/1096  at three different epochs for this particular object.

\begin{table*}
\caption{Spectroscopic data used in this study}
\label{tab:data-source} 
\renewcommand{\arraystretch}{1.2} 
\begin{tabular}{ccccccccc}
\hline
\hline
Object     & Telescope  & Instrument & grating   & $\lambda$ & R     & Date       & Exposure  & PI \\
           &            &            & & (\AA)     &       & (yy/mm/dd) & (s) &     \\ 
\hline
2dFS\,163 &HST & COS & G130M  & 1132--1433 & 12000--17000 & 2020/10/09  &2084  &  Roman-Duval, J  \\
  &HST & COS & G160M  & 1419--1790 & 13000--20000 & 2020/10/09  & 4815 & Roman-Duval, J  \\
  &VLT& X-shooter & UBV  & 3100--5600 &6700 & 2020/10/28  & $2\times1655$ &  Vink, J \\
   &VLT& X-shooter & VIS  & 5600--10240 &11400 & 2020/10/28  &$2\times1725$  &  Vink, J  \\
     &AAT & 2dF &    & 3900--4900 & 1600 & 1998/09/25  & 1800 & Evans, C.J.  \\
\hline

2dFS\,2553 &HST & COS & G130M  & 900--1236 & 13000 & 2022/06/14  &2907  &  Roman-Duval, J  \\
&HST & COS & G130M  & 900--1236 & 13000 & 2022/07/01  &6963 &  Roman-Duval, J  \\
&HST & COS & G130M  & 900--1236 & 13000 & 2022/05/19  &1828  &  Roman-Duval, J  \\

  &HST & STIS & E140M  & 1144--1710  & 45800 & 2020/06/24  & 2755 & Mahy, L  \\
  &VLT& X-shooter & UBV  & 3100--5600 &6700 & 2020/11/08  & $3\times1655$ &   Vink, J \\
   &VLT& X-shooter & VIS  & 5600--10240 &11400 & 2020/11/08  &$2\times1725$  &   Vink, J \\
  &VLT& X-shooter & UBV  & 3100--5600 &6700 & 2022/08/14  & 2000 &  Pauli, D \\  
    &VLT& X-shooter & VIS  & 5600--10240 & 8935 & 2022/08/14  & 2000 &  Pauli, D \\
     &AAT & 2dF &    & 3900--4900 & 1600 & 1998/09/28  &1800 &  Evans, C.J. \\
\hline
 
SK\,-71\,35 &HST & COS & G130M  & 1132--1433 &12000--17000 & 2022/04/20  & 488  &  Roman-Duval, J  \\
  &HST & COS & G160M  & 1419--1790 & 13000--20000 & 2022/04/20  & 816 & Roman-Duval, J  \\
  &VLT& X-shooter & UBV  & 3100--5600 &6700 & 2021/03/12  & 540 &  Vink, J \\
   &VLT& X-shooter & VIS  & 5600--10240 &11400 & 2021/03/12  &610  &  Vink, J  \\ 
     &VLT & GIRAFFE &  LR02  & 3960--4570  & 6300 & 2015/12/19   &  5400 &  Oskinova, L.M  \\
     &VLT & GIRAFFE &  LR03  & 4500-5070  & 7500 & 2015/12/19   &  5400 &  Oskinova, L.M  \\
     &VLT & GIRAFFE &  LR03  & 6445--6820  & 17000 & 2015/12/19   &  3600 &  Oskinova, L.M  \\
\hline

\end{tabular}
\end{table*}

Optical spectra of each star were obtained with  X-shooter at the ESO's Very Large Telescope (VLT). Data reduction and normalization methods are described by Sana et al. (eDR1, submitted). 
The spectral coverage includes both the UBV  and VIS arms. 
In addition, low-resolution archival optical spectroscopy is available for two SMC targets as a part of the 2dF SMC survey \citep[see][for more details]{Evans2004}. VLT/FLAMES spectra of \mbox{Sk\,-71$^{\circ}$\,35} are also utilized in this study \citep[details on the observations and data reduction are in][]{Ramachandran2018,Ramachandran2018b}.
Here, the spectra are taken with three of the standard settings of the Giraffe spectrograph: LR02, LR03, and HR15N. Details of the available spectra for each source are summarized in Table\,\ref{tab:data-source}.

In addition to the flux-calibrated UV spectra, we used various photometric data (from UV to infrared) to construct the spectral energy distribution (SED). The UBV and JHK magnitudes of all sources are taken from Table B.1 in \cite{Vink2023}. Additionally, we included Gaia G magnitudes from \cite{GaiaCollaboration2022}.  
We also utilized data from the Transiting Exoplanet Survey Satellite (TESS) for \mbox{Sk\,-71$^{\circ}$\,35}. We extracted the light curves and used them to understand the binary period of the system. The light curves of the two remaining stars are contaminated by bright sources visible in their full frame images (FFIs) since they are fainter in optical (by 2\,mag).  Therefore, we did not utilize them to determine periods.

\section{Analysis}
\label{sect:analysis}

\subsection{Atmosphere models}
\label{sect:models}
We performed the spectral analysis of the binaries discussed in this paper using the PoWR model atmosphere code.
PoWR is a state-of-the-art stellar atmosphere code suitable for the spectroscopic analysis of hot stars with and without winds, across a broad range of metallicities \citep{Hainich2014,Hainich2015,Oskinova2011}.  The PoWR code solves the radiative transfer equation for a spherically expanding atmosphere and the statistical equilibrium equations simultaneously under the constraint of energy conservation. Stellar parameters were determined iteratively.  Details of the PoWR code are described in \citet{Graefener2002}, \citet{Hamann2003}, \citet{Hamann2004}, \citet{Sander2015}, and \citet{Todt2015}. PoWR models have been used to analyze a large sample of WR stars \citep[e.g.,][]{Hainich2015,Hainich2014,Shenar2016,shenar_2019_wolfrayet}, OB stars \citep[e.g.,][]{Ramachandran2018b,Ramachandran2019,Rickard2022}, and a few stripped stars in binaries \citep{Ramachandran2023,Pauli2022b,Rickard2023} in the Magellanic Clouds. In the discussions, we compare the results from the analysis of our newly discovered stripped star binaries to the above literature as well as other studies. 

The main parameters that specify a PoWR model are the stellar luminosity $L$, stellar temperature $T_\ast$, and surface gravity $g_\ast$, as well as mass-loss rate $\dot{M}$ and terminal velocity $\varv_\infty$ for the wind. \update{The stellar temperature relates to stellar radius $R_\ast$ and $L$ via the Stefan-Boltzmann law $L = 4 \pi \sigma_{\mathrm{SB}}\, R_\ast^2\, T_\ast^4$. In this case the ``stellar temperature'' $T_\ast$ is the effective temperature $T_\mathrm{eff}$ corresponding to the stellar radius $R_\ast$ that marks the inner boundary of the model atmosphere. In our models, the inner boundary and thus $R_\ast$ is located at a Rosseland continuum optical depth of $20$.} Contrary to OB stars, stripped stars may have a slight difference between their stellar temperature $T_\ast$ and the photospheric effective temperature $T_{2/3}$, defined correspondingly to $R_{2/3}$ at $\tau_\mathrm{Ross} = 2/3$. In the derived parameters of our sample, we provide $T_{2/3}$, $\log g_{2/3}$, and $R_{2/3}$, as well as $T_\ast$, $\log g_\ast$, and $R_\ast$. 

In the subsonic region of the stellar atmosphere, a velocity field is defined such that a hydrostatic density stratification is approached \citep{Sander2015}. In the supersonic region, the prespecified wind velocity field $\varv(r)$ is assumed to follow 
the so-called $\beta$-law \citep{CAK1975} 
\begin{equation}
\varv(r) = \varv_\infty \left( 1- \frac{r_{0}}{r} \right)^{\beta}.
\end{equation}
In this work, we adopt $\beta=0.8$, for OB-type secondaries \citep{Kudritzki1989} and we explore different values for stripped stars.

In the non-LTE iteration in the co-moving frame, the line opacity and emissivity profiles are treated as Gaussians with a width following from a constant Doppler velocity $\varv_{\mathrm{Dop}}=30\,\mathrm{km\,s^{-1}}$. In the formal integral for the calculation of the emergent spectrum, the Doppler velocity is split into the depth-dependent thermal velocity and a ``microturbulence velocity'' $\xi(r)$.  We assume $\xi(r) = \rm max(\xi_{min},\, 0.1\varv(\emph{r}))$ for the models, where the photospheric microturbulent velocity $\rm \xi_{min}= 14 $\,km\,s$^{-1} $ is fixed for OB-type secondaries \citep{Hainich2019} and varied as a fit parameter for stripped stars. The pressure broadening is also taken into account in the formal integral.

Optically thin inhomogeneities in the model iterations are described by the  ``clumping factor''  $D$ by which the density in the clumps is enhanced compared 
 to a homogeneous wind of the same $\dot{M}$ \citep{HK98}. For all stars in our study, we account for depth-dependent clumping assuming that clumping begins at the sonic point, increases outward, and reaches a density contrast of $D$ at a radius of $R_{\mathrm{D}}$ \citep{Runacres2002}. If necessary, we adjusted the values of $D$ and $R_{\mathrm{D}}$ for stripped stars, otherwise we set them to $D=10$ and $R_{\mathrm{D}} = \, 10\,R_\ast$. 
 
The PoWR code can account for additional ionization due to X-rays. To match the observations (especially \ion{N}{V} lines in the UV), X-rays were integrated into the model for certain objects in this study. The X-ray emission is modeled using the \cite{Baum1992} approach, assuming only free--free transitions contribute to the flux. \update{A more recent description of the inclusion of X-rays in the PoWR is given in \citet{Sander2018}. Comparisons to CMFGEN and FASTWIND are provided in appendix D of \citet{Bernini-Peron+2023} and appendix B.4 of Sander et al. (2024, submitted, XShootU IV), respectively.}
Since the current generation of PoWR models is limited to spherical symmetry, the X-rays are assumed to arise from an optically thin spherical shell around the star.
The X-ray emission is specified by three free parameters: fiducial temperature of the X-ray emitting plasma $T_{\rm X}$, the onset radius of the X-ray emission $R_0$ $(R_0 > R_\ast)$, and a filling factor $X_{\rm fill}$.  
We set the onset radius to 1.1$R_\ast$ and $T_{\rm X}$ to 1\,MK. The X-ray filling factor is adjusted such that the UV lines are reproduced by the model. 

The models are calculated using complex atomic data of H, He, C, N, O, Mg, Si, P, S, and the Fe group elements. The iron group elements are treated with the so-called superlevel approach as described in \citet{Graefener2002}. The initial chemical abundances of C, N, O, Mg, Si, and Fe are adopted from \cite{Trundle2007} and \cite{Hunter2007}. The abundance of the remaining elements in LMC stars is fixed at $1/2\,Z_\odot$, while in SMC stars it is fixed at $1/5\,Z_\odot$. The mass fractions of carbon (C), nitrogen (N), and oxygen (O) are adjusted as needed for each object to achieve the best possible match with their observed spectra.

The PoWR code is used to calculate models for individual stars. To analyze composite spectra originating from binaries, we compute individual models tailored to each component of the binary and merge them (based on their luminosity ratio) such that we match the composite observed spectra. 
Our models are limited to spherical symmetry, which may not hold for some binary systems. In particular for the case of fast-rotating Oe/Be secondaries, rotation modifies the shape of the star and leads to a deviation from spherical symmetry, thereby influencing the derived stellar parameters such as effective gravity, effective temperature, and radiative flux \citep{vonZeipel1924}. The pole-to-equator temperature and gravity can differ by a few kK and few dex, respectively, resulting in an uncertainty of about $\approx$2 kK in $T_\ast$ and $\approx$0.2 dex in $\log g$ for early main sequence B stars that are close to critically rotating $\Omega/\Omega_{C}=0.8$ \citep{Fremat2005}.

\subsection{2dFS\,163}
\label{2dFS163}
\begin{figure*}  
\vspace{0.5cm}
\includegraphics[scale=0.9]{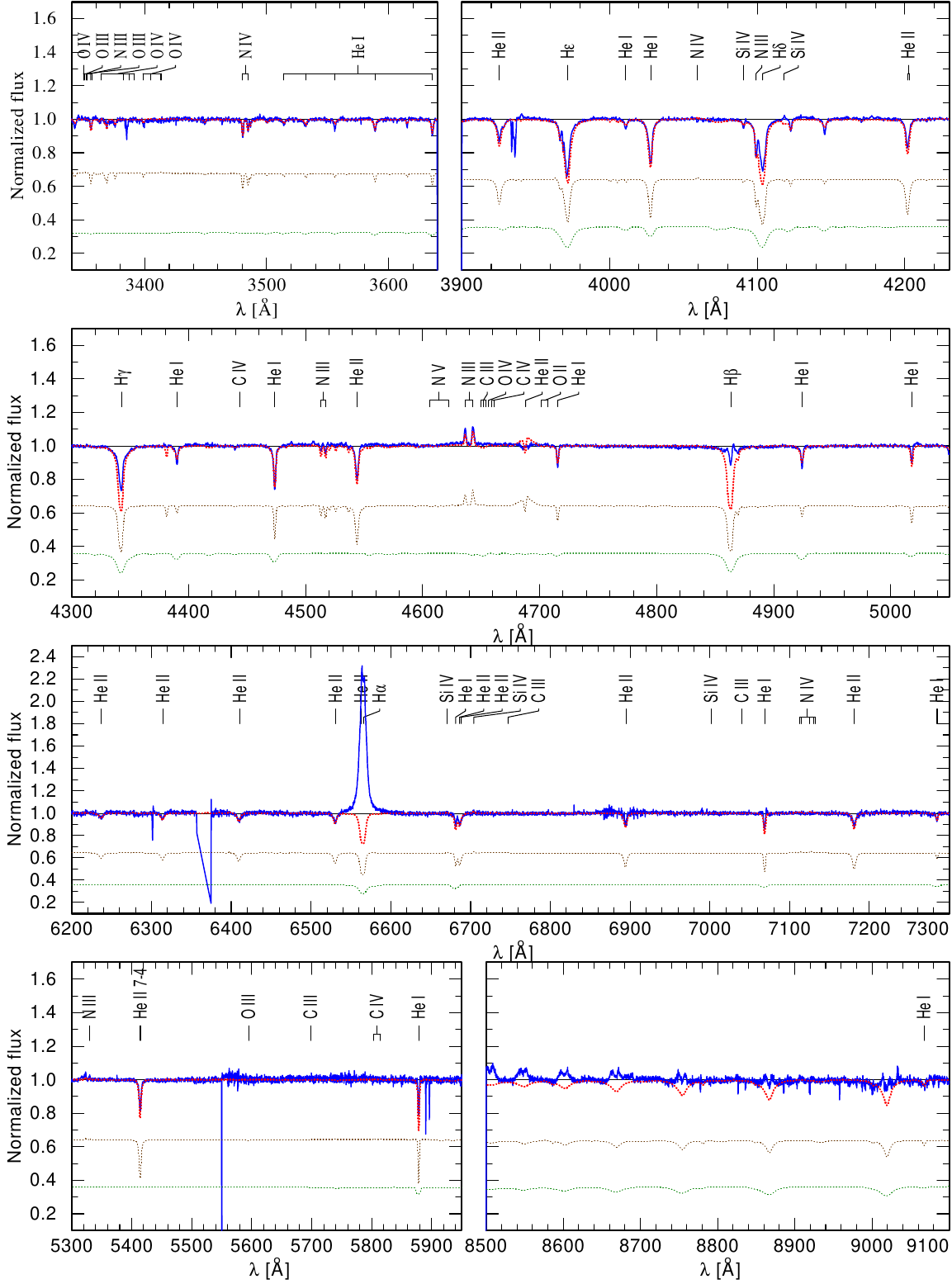}
\caption{Observed spectra of  2dFS\,163 (solid blue) compared to the model spectra.  The composite
model (dotted red) is the weighted sum of the stripped star primary (dotted brown)
and rapidly rotating B star secondary (dotted green) model spectra}
\label{fig:optfit}
\end{figure*}

2dFS\,163 was previously classified as an O8 supergiant by \cite{Evans2004} as a part of 2dF SMC survey. 
The object's optical and UV spectra are primarily characterized by the prevalence of narrow absorption lines, with the exception of weak broad components observed in the \ion{He}{i} and Balmer lines (see Fig.\,\ref{fig:optfit}).
\update{ The optical spectrum can be classified as O7.5 based on the intensities of \ion{He}{i\,$\lambda$4471} and \ion{He}{ii\,$\lambda$4542}, using the O type standards from \cite{Sota2011}. Additionally, the luminosity class can be determined as Ib(f) by considering the intensity of \ion{He}{ii\,$\lambda$4686}, as suggested by \cite{Evans2004}. When solely focusing on the narrow He lines, the intensity ratio between \ion{He}{i\,$\lambda$4471} to \ion{He}{ii\,$\lambda$4542} suggests a slightly earlier subtype than O7, in line with the strength of \ion{N}{iii\,$\lambda\lambda$4634-4640} emission lines. Hence we classified the primary as O6.5\,Ib(f). 
We determine the radial velocity (RV) at different epochs as described in Sect.\,\ref{appendix:2dfs163}. The narrow absorption lines show an RV variation of $\sim28$\,km\,s$^{-1}$ between the recently obtained X-shooter spectra and archival 2dF spectra.
In addition, the narrow photospheric lines in the UV spectra (e.g., \ion{N}{iii}, \ion{Si}{v}) show an RV variation of $\sim$$43$\,km\,s$^{-1}$ relative to X-shooter spectra, indicating its binary nature.
Furthermore, the observed RV variations indicate that the narrow-line star moves faster ($\delta RV \sim 43$\,km\,s$^{-1}$) than the less prominent broad-line star ($\delta RV \sim 15$\,km\,s$^{-1}$), indicating that the O supergiant-like star (partially stripped) likely possesses a significantly lower mass than its rapidly rotating companion.} Moreover, double peak disk emission profiles are visible in H$\alpha$,  H$\beta$, and across all members of the Paschen series.
These lines are likely coming from the fast-rotating companion rather than the slow-rotating O supergiant-like star. In conclusion, the spectral features suggest that 2dFS\,163 is a post-interaction binary system consisting of a star that has been stripped of its outer layers and rapidly rotating early Be star companion.

\begin{figure*}  
\vspace{0.5cm}
\includegraphics[scale=0.9,trim={0cm 0cm 0cm 6.5cm},clip]{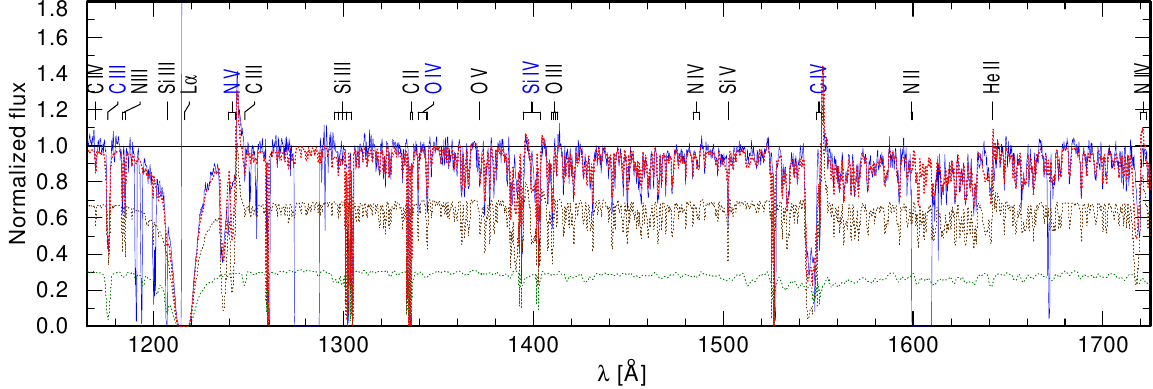}
\caption{Observed HST UV spectra of 2dFS\,163 (solid blue) compared to the model spectra.  The composite
model (dotted red) is the weighted sum of the stripped star primary (dotted brown)
and rapidly rotating B star secondary (dotted green) model spectra}
\label{fig:uvfit}
\end{figure*}

We chose PoWR SMC grid models \citep{Hainich2019} as a starting point for the analysis and further computed additional models with tailored parameters. As an initial approach, we try to fit the spectra as a single star since the primary contributes to most of the light. One can approximately fit the spectra with models assuming a range of stellar temperatures $T_{\ast} \sim 34-38$\,kK. For the cooler models, we used a lower $\log\,g_\ast$ (3.5) while a higher $\log\,g_\ast$ (3.9) is used for the hotter models. The cooler models were unable to reproduce the \ion{N}{iii} emission lines and the \ion{N}{iv} absorption lines in the optical as well as the unsaturated \ion{C}{iv} and weak \ion{He}{ii} line in the UV. On the other hand, the hotter models result in weaker \ion{He}{i} lines and broader \ion{He}{ii} lines in the optical as well as weaker \ion{C}{iv} and stronger \ion{He}{ii} in the UV than observed. Despite the differences, all these single-star models suggested much lower stellar masses than expected for an O supergiant, namely in the range of $\sim 7-12 M_{\odot}$.

After these initial considerations, we proceed with a two-model approach representing each of the components. To determine the temperature of the stripped primary star, we utilized the \ion{N}{iii\,$\lambda\lambda$4634--4642} emission lines and various absorption lines including \ion{N}{iv\,$\lambda\lambda$3460-3485}, \ion{Si}{iv\,$\lambda$4089}, and \ion{He}{ii} lines in the optical spectrum. These lines are exclusively originating from the stripped primary star, making them suitable for constraining its temperature. Simultaneously, the \ion{C}{iv\,$\lambda$1169} to \ion{C}{iii\,$\lambda$1175} and \ion{O}{iv} to \ion{O}{v} line ratios are employed in the ultraviolet spectra to constrain the temperature of the primary. 

In order to account for the influence of gravity on the ionization balance, we made simultaneous adjustments to both the effective temperature ($T_\ast$) and the surface gravity ($\log\,g_\ast$) to obtain a satisfactory agreement with the observed spectra. 
A major constraint on  $\log\,g_\ast$ of the primary is the need to align with the profiles of the \ion{He}{ii} spectral lines. The Balmer lines are not a good diagnostic in this case as they are influenced by contributions from both the secondary and primary star as well as the disk contribution. Along with $T_\ast$ and $\log\,g_\ast$, we also varied the micro-turbulence $\xi$ velocity in the range of 10-20\,km\,s$^{-1}$ to match the metal line profiles.

The secondary star is not making any significant contribution to the metal lines or the \ion{He}{ii} lines in the optical (no broad components visible in the spectra), suggesting that its temperature is unlikely to exceed 30,000\,K, as otherwise there would be observable contributions in the \ion{He}{ii} lines. Similarly, there should be \ion{Si}{iii} lines in the optical near 4550-4570\,\AA\ if the secondary would be cooler than 24,000\,K. Simultaneously, the secondary component exhibits noticeable effects on the absorption lines of \ion{He}{i}, specifically at wavelengths 4387, 4471, and 4922 \AA.  Based on these, we could estimate the temperature range and subsequently make adjustments to the luminosity ratio.
The luminosity ratios, surface gravities, and temperatures in the primary and secondary models were adjusted concurrently to match the observations. 

The projected rotation velocity ($\varv\,\sin i$) of the narrow-lined primary was estimated using \ion{N}{iii} emission lines, as well as absorption lines of \ion{N}{iii}, \ion{N}{iv}, and \ion{He}{i}. In our study, we employed a combined approach of Fourier transform (FT) and goodness-of-fit (GOF) analysis using the \texttt{iacob-broad} tool \citep{Simon-diaz2014}. The method was applied to various lines and the average values are adopted for $\varv\sin i$ and the macro-turbulent ($\varv_{\mathrm{mac}}$) velocity. In the secondary analysis, a high rotational velocity was applied to the model in order to accommodate the broad core of the \ion{He}{i} lines. In this case, we fixed  $\varv_{\mathrm{mac}}$ while varying $\varv\sin i$ to achieve the best fit of these lines. These velocities were then used, along with instrumental broadening, to convolve the model spectra such that they match the observations.

We initially computed models with typical SMC abundances \citep{Trundle2007}. However, they fail to reproduce all the observed CNO lines satisfactorily. In order to match the observed intensity of the CNO lines, the abundance of N is increased by a factor of 50 in the stripped-star model, while the abundances of C and O are decreased by a factor of 20 and 5, respectively. The N abundance is determined by analyzing multiple \ion{N}{iv} and \ion{N}{iii} lines in both the UV and the optical range. The optical range exhibits a notable absence of the majority of C and O absorption lines (see Fig.\,\ref{fig:optfit}) which is usually not expected in an O star spectrum. This absence suggests a strong depletion in the abundance of these elements.
For the C abundance, we rely on the \ion{C}{iv\,$\lambda$1169} and \ion{C}{iii\,$\lambda$1175} absorption lines in the UV. We deduce the O abundance from \ion{O}{iv} and \ion{O}{iii} lines near 3350\AA\ and  \ion{O}{iv} lines in the UV.
\update {In addition to the CNO abundances, we varied the H mass fraction ($X_{\rm H}$) in the primary model between 0.2 and 0.73 and found that H-depleted (He-enriched) models better represent the observations. At this hot temperature, the strength of narrow components in \ion{He}{i} and \ion{He}{ii} lines can only be simultaneously reproduced with a high He abundance. A standard He abundance would result in very weak or absent \ion{He}{i} lines at 4388, 4713, 4922 \AA\ for the stripped star. Additionally, in the UV spectrum, 
\ion{He}{ii\,$\lambda$1640} line is found to be better reproduced with this higher He abundance. } Strong He and N enrichment, along with C and O depletion, are in agreement with the primary's stripped envelope.

\begin{figure}[!htb]  
\vspace{0.5cm}
\includegraphics[scale=0.35]{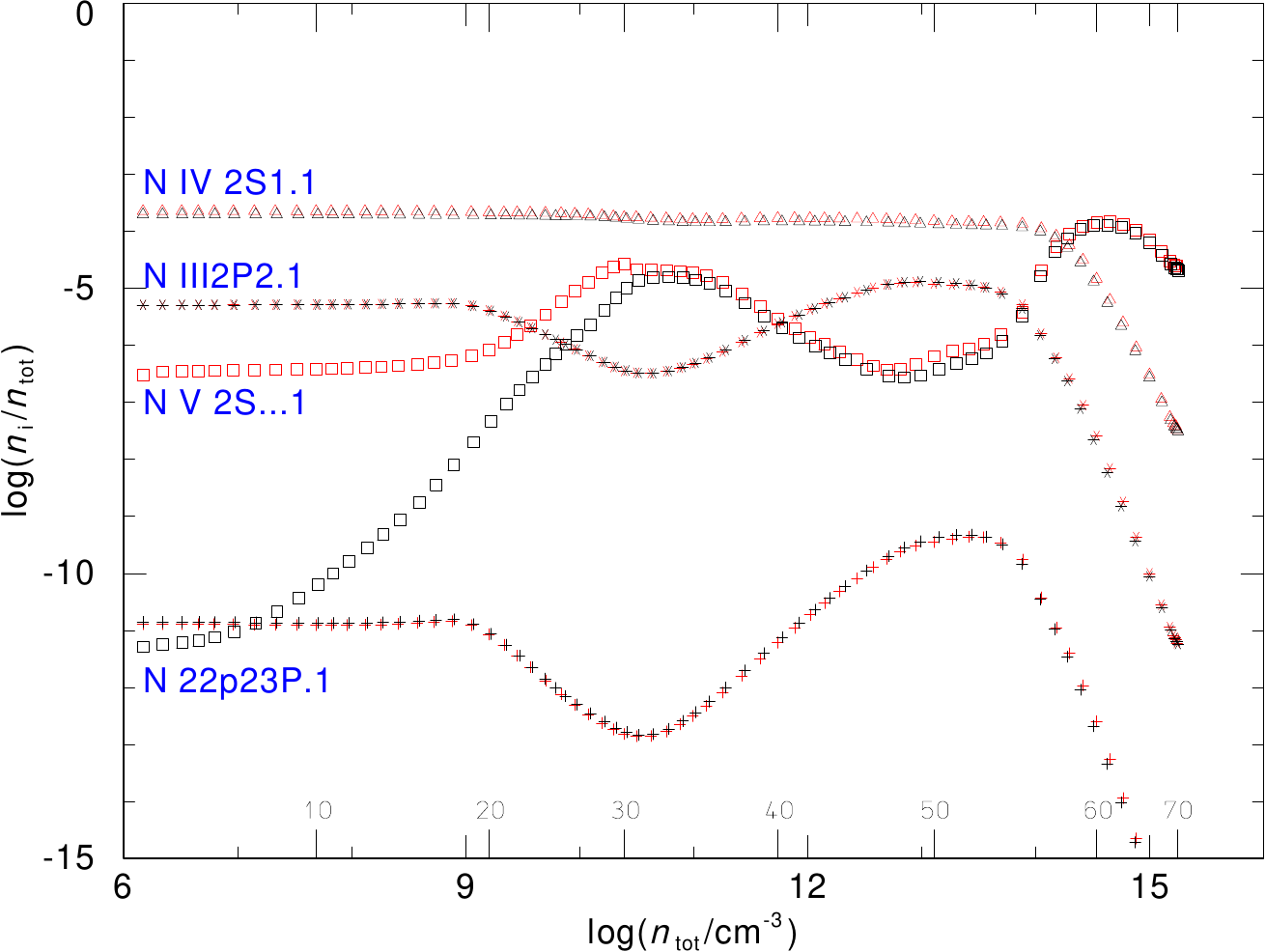}
\caption{Change in nitrogen level population in the model when including X-rays (red) and without X-rays (black). The levels of nitrogen ions \ion{N}{ii}, \ion{N}{iii}, \ion{N}{iv}, \ion{N}{v}  are denoted by a plus sign, asterisk, triangle, and square, respectively. }\label{fig:Nlevel}
\end{figure}  

\begin{figure}[!htb]  
\vspace{0.5cm}
\includegraphics[scale=0.88,trim={0cm 6.7cm 0cm 0cm},clip]{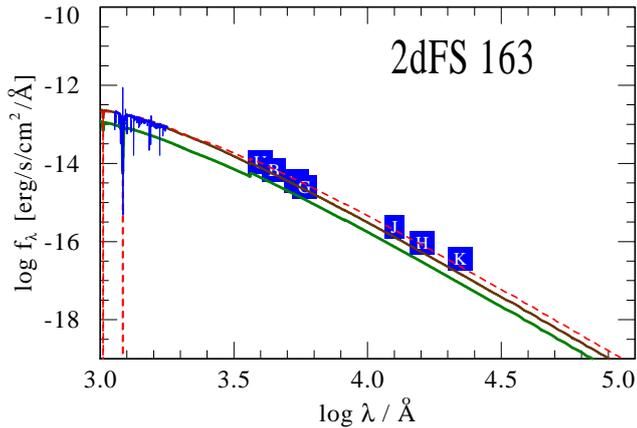}
\caption{Observed flux calibrated UV spectra and photometry of 2dFS\,163 (blue) compared to the model SED.  The composite
model (dashed red) is the weighted sum of the stripped star primary (brown) and rapidly rotating B star secondary (green) model}
\label{fig:sedfit}
\end{figure}

\begin{table}[!hbt]
	\caption{Fundamental parameters and abundances derived for 2dFS\,163 using spectroscopic analysis.}
	\label{table:parameters}
	\centering
	\renewcommand{\arraystretch}{1.6}
	\begin{tabular}{lcc}
		\hline 
		\hline
		\vspace{0.1cm}
&	Stripped star                                     &     Be star           \\
		\hline 
  Spectral type & O6.5\,Ib (f)  & B0-1 Ve\\
		$T_{\ast}$ (kK)                                & $37^{+1}_{-2}$       &  $26^{+3}_{-4}$ \\
 $T_{2/3}$ (kK)                                 & $35.9^{+1}_{-2}$ &  $25.8^{+3}_{-4}$ \\
   
		$\log g_\ast$ (cm\,s$^{-2}$)    & $3.5^{+0.1}_{-0.2}$   &   $3.8^{+0.2}_{-0.2}$   \\
 $\log g_{2/3}$ (cm\,s$^{-2}$)    & $3.45^{+0.1}_{-0.2}$    &    $3.79^{+0.2}_{-0.2}$   \\
 $\log g_\mathrm{true}$\tablefootmark{\dag} (cm\,s$^{-2}$)    & $3.51^{+0.1}_{-0.2}$   &   $3.89^{+0.2}_{-0.2}$   \\
		$\log L$ ($L_\odot$)                           & $4.75^{+0.1}_{-0.1}$  &   $4.18^{+0.2}_{-0.2}$    \\ 
		$R_\ast$ ($R_\odot$)                           & $5.8^{+1}_{-0.7}$    &    $6.1^{+2.5}_{-2}$      \\
  $R_{2/3}$  ($R_\odot$)                               & $6.1^{+1}_{-0.7}$  &   $6.15^{+2.5}_{-2}$    \\
		$\log \dot{M}$ ($M_\odot \mathrm{yr}^{-1}$)    & $-6.9^{+0.1}_{-0.2}$  &-10 (fixed)  \\
		$\varv_{\infty}$ (km\,s$^{-1}$)     & $1000^{+100}_{-100}$ & 1500 (fixed)  \\ 
$\beta$                               &   1.1          & 0.8  (fixed)      \\
		$D$                               &   20          & $10$  (fixed)      \\
		$\varv \sin i$ (km\,s$^{-1}$)       & $60^{+10}_{-10}$   &  $250^{+80}_{-80}$      \\
		$\varv_{\mathrm{mac}}$ (km\,s$^{-1}$)         & $50^{+10}_{-10}$   &    50 (fixed)   \\
  $\varv / \varv_{\mathrm {crit}}$       & $\gtrsim 0.2$   &  $\gtrsim 0.53$      \\
		$\xi$ (km\,s$^{-1}$)            & $15^{+5}_{-5}$     &      $14$  (fixed)   \\
		$X_{\rm H}$ (mass fr.)                         & $0.33^{+0.1}_{-0.05}$  &   0.737\tablefootmark{$\ast$}  \\
		$X_{\rm He}$ (mass fr.)                         & $0.67^{+0.05}_{-0.1}$  &   0.26\tablefootmark{$\ast$}   \\
		$X_{\rm C}/10^{-5}$ (mass fr.)                 & $1^{+1}_{-0.5}$       &   21\tablefootmark{$\ast$}     \\
		$X_{\rm N}/10^{-5}$ (mass fr.)                 & $160^{+20}_{-20}$  & 3\tablefootmark{$\ast$}          \\
		$X_{\rm O}/10^{-5}$ (mass fr.)                 & $20^{+10}_{-10}$      &   113\tablefootmark{$\ast$}      \\
		$E_{\mathrm{B-V}}$ (mag)                                & $0.08^{+0.02}_{-0.02}$ &    \\
		$M_\mathrm{spec}$ ($M_\odot$)                  & $3.96^{+1.8}_{-1.5}$    &  $10.5^{+5}_{-4}$      \\
		$\log\,Q_{\mathrm H}$ (s$^{-1}$)   &48.4  &  46.4   \\
		$\log\,Q_{\mathrm {He\,\textsc{ii}}}$ (s$^{-1}$)   & 39.8 &  34.7   \\
		\hline
	\end{tabular}
	\tablefoot{ \tablefoottext{$\dag$} {$\log (g_\ast +(\varv \sin i)^{2} /R_\ast)$}
 \tablefoottext{$\ast$} {Abundances of the Be star are adopted from \citet{Trundle2007} which correspond to typical values for OB stars in the SMC}
 }
 \vspace{-0.2cm}
\end{table}

Once the stellar parameters are adjusted, we proceed to modify the wind parameters in the primary model in order to accurately match the UV P\,Cygni profiles, and wind lines observed in the optical spectrum. 
We measured $\varv_\infty$ from the blue edge of \ion{C}{iv\,$\lambda\lambda$1548--1551} and \ion{N} {v\,$\lambda\lambda$1238--1242} (see Fig.\,\ref{fig:uvfit}). The mass-loss rate ($\dot{M}$) is estimated from the \ion{C}{iv}, \ion{N}{v} and \ion{Si}{iv} profiles. 
Nevertheless, an accurate reproduction of \ion{N}{v} can only be achieved by incorporating an X-ray field into our model, where the X-ray luminosity is $\log L_{\rm X} =31.5$\,erg\,s$^{-1}$. X-rays have a significant impact on the ionization structure, particularly in enhancing the \ion{N}{v} line \citep{Cassinelli1979, Baum1992}. Figure\,\ref{fig:Nlevel} demonstrates the influence of incorporating X-rays on the population of nitrogen ions. Specifically, the incorporation of X-rays leads to an increase in the population of \ion{N}{v} in the outer wind. It should be noted that the sensitivity of existing X-ray surveys is not sufficient to detect the star. Using current
X-ray catalogs, the upper limit on X-ray luminosity of 2dFS\,163 in the 0.2-12.0 keV band is $\log{L_X}<36.5$. 

 Along with $\dot{M}$, we also adjusted the wind velocity law exponent. Assuming  $\beta$ = 1.1 is found to be in best agreement with the observed P\,Cygni profiles. Because clumping enhances the emission in optical wind lines such as H$\alpha$ and \ion{He}{ii\,$\lambda$4686}, we varied the clumping factor and the onset of clumping to best match these lines. In this system, since H$\alpha$ is filled with disk emission from the companion, we had to solely rely on \ion{He}{ii\,$\lambda$4686}, which nearly matches the continuum due to filled-in emission. This line is best reproduced by a model with a density contrast of $D=20$ where the clumping starts at a radius of 1.12$\,R_\ast$ and reaches the maximum value of $D$ at $R_{\mathrm{D}} = 5\,R_\ast$. Given the limited contribution of the secondary component to the UV lines and the relatively weak wind of B dwarfs in the SMC, we assumed a fixed low $\dot{M}$ and other wind parameters for the model. The only impact of the secondary star on the composite UV spectra is the dilution of lines from the primary. One such noticeable feature is the unsaturated \ion{C}{iv} P\,Cygni profile with a flat bottom. \update{The non-zero flux in the absorption trough of the \ion{C}{iv\,$\lambda\lambda$1548--1551}  line
suggests that the stripped star wind does not substantially occult the
secondary. Thus, the binary system may be widely separated.}

 \update{ We determine the luminosity $L$ and color excess $E_{\rm B-V}$ by fitting the composite model SED to the photometry and flux-calibrated UV spectra (Fig.\,\ref{fig:sedfit}). The model flux is diluted with the adopted SMC distance modulus of 18.98\,mag \citep{Graczyk2020}. The reddening encompassed contributions from an internal SMC component and the Galactic foreground. For the
Galactic component we used the reddening law published by \cite{Seaton1979}
and a color excess of $E_{\rm B-V}$= 0.04 mag. For the SMC component, the color excess is varied as a free parameter, while the reddening law is adopted from \cite{Bouchet1985}.}
By fitting the normalized spectra, we can place constraints on the luminosity ratio, thus the SED fitting by the composite model yields both primary and secondary luminosities. For 2dFS 163, the primary stripped star contributes approximately 70\%  of the light in the UV and 65\% in the optical. 

The final, best-fit composite models for 2dFS\,163 are shown in Figs.\,\ref{fig:optfit}, \ref{fig:uvfit}, and \ref{fig:sedfit} respectively. 
 The resulting stellar and wind parameters as well as the derived surface abundances for both primary and secondary components are given in Table\,\ref{table:parameters}.

\begin{figure}
    \centering
    \includegraphics[width=0.9\linewidth]{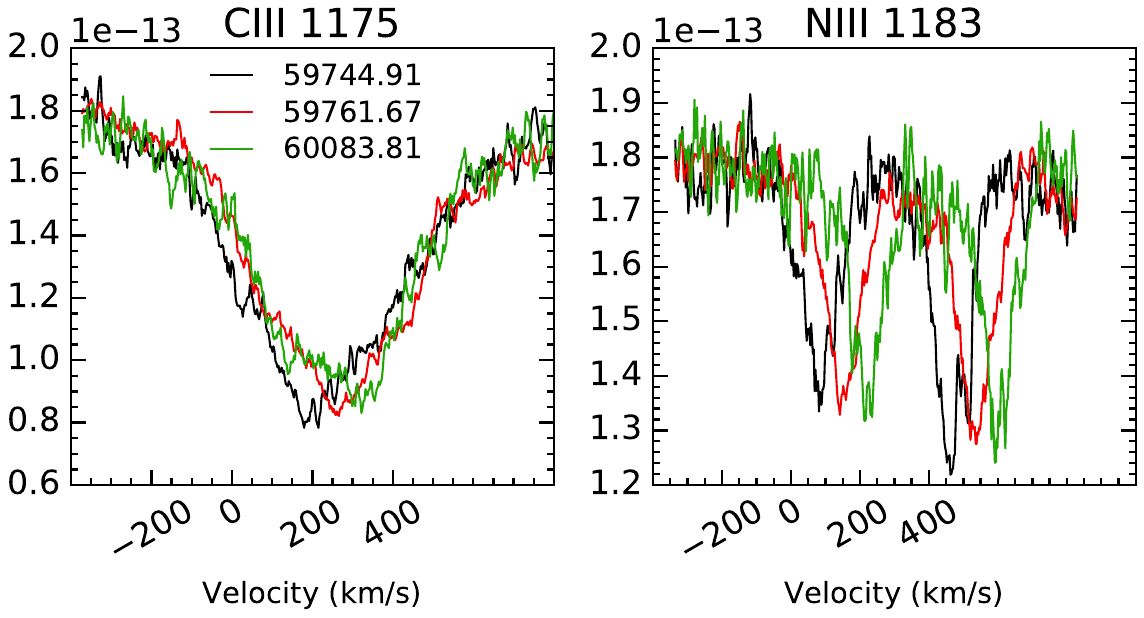}
    \caption{UV spectra of 2dFS\,2553 taken at different epochs showing radial velocity shift in \ion{C}{iii\,$\lambda$1175} and \ion{N}{iii\,$\lambda$1183,1185} lines. }
    \label{fig:RVUV2553}
\end{figure}

\begin{figure}
    \centering
    \includegraphics[width=0.9\linewidth]{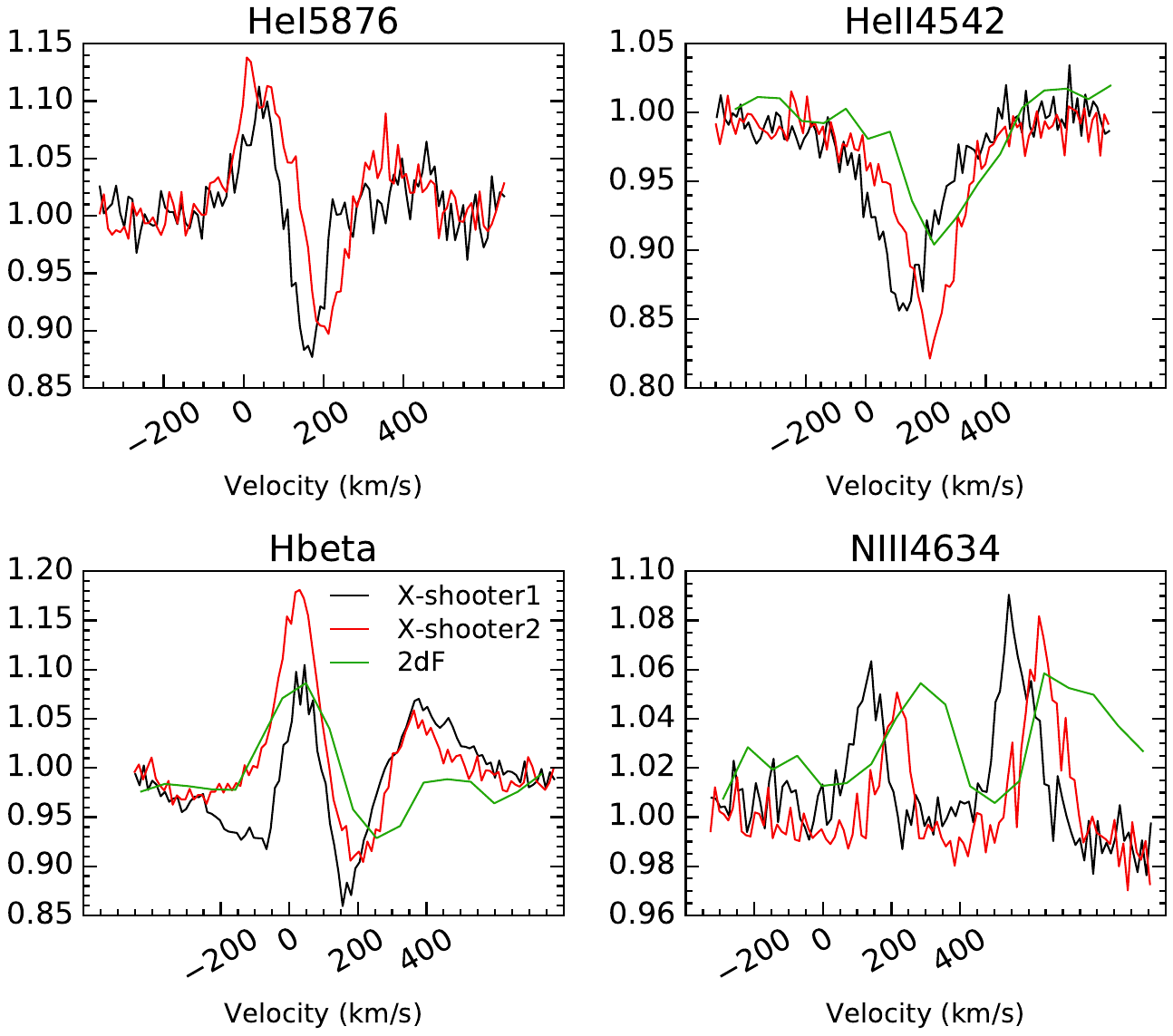}
    \caption{Radial velocity shift between 2dF and two X-shooter spectra of 2dFS\,2553. The narrow primary lines are moving in opposite directions to the broad and disk emission components.}
    \label{fig:RV2553}
\end{figure}

\subsection{2dFS\,2553}
\label{2dFS2553}

\cite{Evans2004} classified the star as O6.5II(f) using the 2dFS survey. 
Based on multi-epoch UV and optical spectra, we found that 2dFS\,2553 shows RV variations (see Figs.\,\ref{fig:RVUV2553}, \ref{fig:RV2553}). Both the UV and optical spectra are dominated by narrow absorption and emission lines from the primary star. The \ion{N}{iii\,$\lambda\lambda$4634--4642} and \ion{N}{iv\,$\lambda$4057} emission lines as well as the \ion{N}{iv\,$\lambda\lambda$3460-3485} absorption lines and all \ion{He}{ii} lines stem exclusively from the hot primary. Based \ion{N}{iv} to \ion{N}{iii} ratio, we reclassify the primary as O4\,III\,(f). Narrow lines in the UV spectra such as \ion{N}{iii\,$\lambda$1183} and \ion{C}{iv\,$\lambda$1169} show RVs in the range of $\sim$80-210 km\,s$^{-1}$ (see Fig.\,\ref{fig:RVUV2553}). In the optical, \ion{N}{iii} and \ion{N}{iv} emission lines show  RVs in the range of $\sim$130-260 km\,$^{-1}$ (see Fig.\,\ref{fig:RV2553}). 

The weak broad-line components from the secondary are visible in the X-shooter spectra, mainly in the \ion{He}{i} lines. Specifically, the \ion{He}{i} lines at 4144 and 4388\AA\ are exclusively from the secondary (no narrow component). The secondary also shows disk emission line contribution to the Balmer and Paschen lines as well as to the \ion{He}{i} lines in the red optical part. It is further evident from Fig.\,\ref{fig:RV2553} that the emission part in \ion{He}{i\,$\lambda$5876} and H$\beta$ are moving in the opposite direction compared to the narrow absorption part and the \ion{N}{iii} emission lines. All this suggests that the secondary is a fast-rotating, early-type Be star. 
We measured an RV separation of $\approx90$ km/s between primary and secondary lines in the X-shooter spectra. Between two epochs of X-shooter spectra, the secondary's lines are moving $\approx70$\,km/s in the opposite direction of the primary. The \ion{C}{iii\,$\lambda$1175} line in the UV range also shows strong contributions from the broad components of the secondary, but the RV variations in the line are much smaller than that found in narrow \ion{N}{iii\,$\lambda$1183} lines (see Fig.\,\ref{fig:RVUV2553}). In conclusion, the luminous primary is moving faster than the Be secondary, indicating that the mass of the primary is smaller than that of the secondary star.

Details of the RV measurements are given in appendix Sect.\,\ref{appendix:RV}. 
Since the narrow lines are prominent in both the UV and optical spectra, we measured the RV of the primary from 7 epochs, as listed in the appendix Table \ref{tab:RV}. An additional RV value for the star is reported by \citet{Lamb2016}. Using all this information, we searched for possible orbital periods of the system with \textit{The Joker} \citep{Price-Whelan2017}, a custom Monte Carlo sampler for sparse RV datasets. In addition, we employed PHOEBE to derive the period and to determine orbital parameters. The most likely period of the system is found to be 93.6 days (see more details in appendix Sections\,\ref{appendix:Joker} and \ref{appendix:PHOEBE}). 


The spectral analysis of 2dFS\,2553 is carried out in the same manner as that of 2dFS\,163 since the composite spectra look similar. The main differences include the presence of the \ion{N}{iv\,$\lambda$4057} emission line in the optical range and a stronger \ion{C}{iv\,$\lambda$1169} to \ion{C}{iii\,$\lambda$1175} ratio in the UV. To account for this, we had to increase the temperature of the primary along with increasing $\log\,g_\ast$.  Since the primary is much hotter here, the weak, broad \ion{He}{i} lines are found to be exclusively from the secondary. This allows us to adjust the luminosity ratios in the composite spectral fit. Again, we found a plausible temperature range for the secondary, as it does not contribute to the \ion{He}{ii} or \ion{Si}{III} lines in the optical spectra similar to 2dFS\,163. The final parameters determined in Table\,\ref{table:parameters2} correspond to the best fit-composite model for the observations.

\begin{table}
	\caption{Fundamental parameters and abundances derived for 2dFS\,2553}
	\label{table:parameters2}
	\centering
	\renewcommand{\arraystretch}{1.6}
	\begin{tabular}{lcc}
		\hline 
		\hline
		\vspace{0.1cm}
&	Stripped star                                     &     Be star           \\
		\hline 
    Spectral type & O4 III (f)  & B0-1 Ve\\
		$T_{\ast}$ (kK)                                & $42^{+1}_{-2}$       &  $27^{+3}_{-3}$ \\
 $T_{2/3}$ (kK)                                 & $41.3^{+1}_{-2}$ &  $26.8^{+3}_{-3}$ \\
   
		$\log g_\ast$ (cm\,s$^{-2}$)    & $3.8^{+0.1}_{-0.2}$   &   $3.8^{+0.2}_{-0.2}$   \\
 $\log g_{2/3}$ (cm\,s$^{-2}$)    & $3.77^{+0.1}_{-0.2}$    &    $3.79^{+0.2}_{-0.2}$   \\
  $\log g_\mathrm{true}$\tablefootmark{\dag} (cm\,s$^{-2}$)    & $3.81^{+0.1}_{-0.2}$   &   $3.91^{+0.2}_{-0.2}$   \\
		$\log L$ ($L_\odot$)                           & $4.91^{+0.1}_{-0.1}$  &   $4.34^{+0.2}_{-0.2}$    \\ 
		$R_\ast$ ($R_\odot$)                           & $5.4^{+1}_{-0.7}$    &    $6.8^{+2}_{-1.6}$      \\
  $R_{2/3}$  ($R_\odot$)                               & $5.59^{+1}_{-0.7}$  &   $6.86^{+2}_{-1.6}$    \\
		$\log \dot{M}$ ($M_\odot \mathrm{yr}^{-1}$)    & $-7.3^{+0.2}_{-0.2}$  &-10 (fixed)  \\
		$\varv_{\infty}$ (km\,s$^{-1}$)     & $1350^{+100}_{-100}$ & 1500 (fixed)  \\ 
  $\beta$                               &   1        & 0.8  (fixed)      \\
		$D$                               &   20          & $10$  (fixed)      \\
		$\varv \sin i$ (km\,s$^{-1}$)       & $80^{+10}_{-10}$   &  $300^{+100}_{-50}$      \\
		$\varv_{\mathrm{mac}}$ (km\,s$^{-1}$)         & $30^{+10}_{-10}$   &    50 (fixed)   \\
  $\varv / \varv_{\mathrm {crit}}$       & $\gtrsim 0.2$   &  $\gtrsim 0.59$      \\
		$\xi$ (km\,s$^{-1}$)            & $15^{+5}_{-5}$     &      $14$  (fixed)   \\
		$X_{\rm H}$ (mass fr.)                         & $0.6^{+0.1}_{-0.05}$  &   0.737\tablefootmark{$\ast$}  \\
		$X_{\rm He}$ (mass fr.)                         & $0.39^{+0.05}_{-0.1}$  &   0.26\tablefootmark{$\ast$}   \\
		$X_{\rm C}/10^{-5}$ (mass fr.)                 & $2^{+1}_{-1}$       &   $4^{+2}_{-2}$    \\
		$X_{\rm N}/10^{-5}$ (mass fr.)                 & $200^{+80}_{-40}$  & 3\tablefootmark{$\ast$}          \\
		$X_{\rm O}/10^{-5}$ (mass fr.)                 & $40^{+20}_{-20}$      &   113\tablefootmark{$\ast$}      \\
		$E_{\mathrm{B-V}}$ (mag)                                & $0.10^{+0.02}_{-0.02}$ &    \\
		$M_\mathrm{spec}$ ($M_\odot$)                  & $6.86^{+3.3}_{-2.8}$    &  $13.8^{+5}_{-3.5}$      \\
		$\log\,Q_{\mathrm H}$ (s$^{-1}$)   &48.4  &  46.4   \\
		$\log\,Q_{\mathrm {He\,\textsc{ii}}}$ (s$^{-1}$)   & 43.1 &  35.9   \\
$P_{\mathrm{orb}}$ (days)    & 93.6 \\
$R_\mathrm{RL}\,(R_\odot)$ & $81.2^{+38}_{-30}$ & $99.7^{+47}_{-38}$\\
$f_\mathrm{RL}$ & $0.07^{+0.035}_{-0.03}$ & $0.07^{+0.035}_{-0.03}$\\
		\hline
	\end{tabular}
	\tablefoot{ \tablefoottext{$\dag$} {$\log (g_\ast +(\varv \sin i)^{2} /R_\ast)$}
 \tablefoottext{$\ast$} {Abundances of the Be star are adopted from \citet{Trundle2007} which correspond to typical values for OB stars in the SMC}
 }
\end{table}
 
We varied the CNO abundance in the stripped primary model to accurately replicate the observed intensity of the CNO lines in the optical and UV spectra. The N enrichment is found to be stronger in 2dFS\,2553 than in 2dFS\,163. However, it is difficult to match the strength of the N lines of different ionization stages simultaneously. The emission lines of \ion{N}{iv} and \ion{N}{iii} in the optical spectrum, as well as the absorption lines of \ion{N}{iii}, and the \ion{N}{V} P\,Cygni profile in the ultraviolet spectrum, are in good agreement with the final nitrogen abundance employed in the model. Conversely, the model slightly under-predicts the strength of the \ion{N}{iv\,$\lambda\lambda$3460-3485} absorption lines in the optical. 
Interestingly, we see a complete absence of C and O absorption lines in the optical spectra (see Fig.\,\ref{fig:optfit}), even in the bluest part where we found a few O lines in 2dFS\,163. Hence, we need to rely on UV absorption lines to determine the C and O abundances. Since the secondary also contributes to the \ion{C}{iii\,$\lambda$1175} line, we could constrain the C abundance in the Be star model, which is also found to be significantly depleted compared to the initial value. 
Unlike 2dFS\,163, there are no obvious indications in the observed spectra for He enrichment. We varied the H mass fraction ($X_{\rm H}$) in the primary model between 0.5 and 0.73 and found the best fit at $X_{\rm H}=0.6$.

 
The wind parameters of the stripped star are determined by analyzing the \ion{C}{iv\,$\lambda\lambda$1548--1551} and \ion{N} {v\,$\lambda\lambda$1238--1242} P\,Cygni lines (see Fig.\,\ref{fig:uvfit2}). The $\varv_\infty$ is measured from the blue edge of the profiles, and the mass-loss rate is estimated from the overall strength of the P\,Cygni profiles. Since the stripped star in 2dFS\,2553 is much hotter than 2dFS\,163, we can accurately replicate the \ion{N}{v} line profile without incorporating any X-rays into the model. The default clumping parameters from the SMC OB grid model \citep{Hainich2019} give a reasonable fit to the phosphorus lines at 1118, 1128 \AA\ in the UV. Similar to 2dFS\,163, the contribution of the secondary component to the UV spectra is very low, especially to the wind lines. Hence, we assumed fixed wind parameters for the secondary model as described in Sect.\,\ref{2dFS163}. The only impact of the secondary star on the composite UV spectra is the dilution of the primary's wind lines and its contribution to the \ion{C}{iii} absorption line.

As described in Sect.\,\ref{2dFS163}, the luminosity $L$ and color excess $E_{\rm B-V} $ are estimated by fitting the composite model SED to the photometry and flux calibrated UV spectra (Fig.\,\ref{fig:sedfit2}). In this system, the stripped star contributes approximately 60\% of the light in the UV and the optical.
The final best-fit model and SED for 2dFS\,2553 are shown in Figs.\,\ref{fig:optfit2}, \ref{fig:uvfit2}, and \ref{fig:sedfit2} respectively.
The stellar and wind parameters as well as surface abundances derived for both the stripped star and the Be star are given in Table\,\ref{table:parameters2}.
 Based on the parameters derived from spectroscopic analysis and orbital analysis (see Sect.\,\ref{appendix:2dfs2553}) we calculated the Roche lobes surrounding each component using $$ \frac{R_\mathrm{RL1}}{a}  = \frac{0.49q^{2/3}}{0.6q^{2/3} + \mathrm{ln}(1+q^{1/3})}$$; where $q=M_1/M_2$ \citep{Eggleton1983}. Subsequently, the Roche lobe filling factor $f_\mathrm{RL}= R_\ast/R_\mathrm{RL} $ is calculated.  The radius of both components of 2dFS\,2553 are much smaller than their Roche lobes, indicating it is a detached binary.

\subsection{\mbox{Sk\,-71$^{\circ}$\,35}}
\label{SK-7135}

This LMC star was classified as a B1\,II in \cite{Ramachandran2018b} and analyzed using a single epoch GIRAFFE spectrum. Although the spectra show B supergiant-like features, it also consists of double peak disk emission features similar to the newly identified partially stripped binary SMCSGS-FS\,69 \citep{Ramachandran2023}.
The optical spectra are primarily characterized by narrow absorption lines originating from the B supergiant-like star. Nevertheless, we have detected absorption lines of \ion{He}{ii\,$\lambda$4686} and \ion{He}{ii\,$\lambda$5412} in the spectra. These \ion{He}{ii} lines are significantly broader than the metal lines in the spectra and can only be attributed to a star with a hotter temperature. 
In addition, the existence of disk emission characteristics in the Balmer and Paschen series indicates the presence of a rapidly rotating Oe star as the secondary companion in the system. The primary, on the other hand, is a slowly rotating, partially stripped star with a substantial H-rich envelope. \update{
Following Fitzpatrick's LMC B supergiant scheme the primary supergiant can be classified as B1 based on the relative strengths of \ion{Si}{iv\,$\lambda$4089} to \ion{Si}{iii\,$\lambda\lambda$4552-4575}. The luminosity class is unclear given the narrow core + broad wings of H$\gamma$, nevertheless, we suggest Ia luminosity class for primary following the new B supergiant Milky Way template scheme of Negueruela+ (submitted) assuming the broad component is from secondary.  
Based on the intensity ratio of the broad component of  \ion{He}{i\,$\lambda$4388} to  \ion{He}{ii\,$\lambda$4542} lines and Balmer emissions suggest that the secondary is O9 e star. }

\begin{figure}
    \centering
    \includegraphics[width=0.9\linewidth]{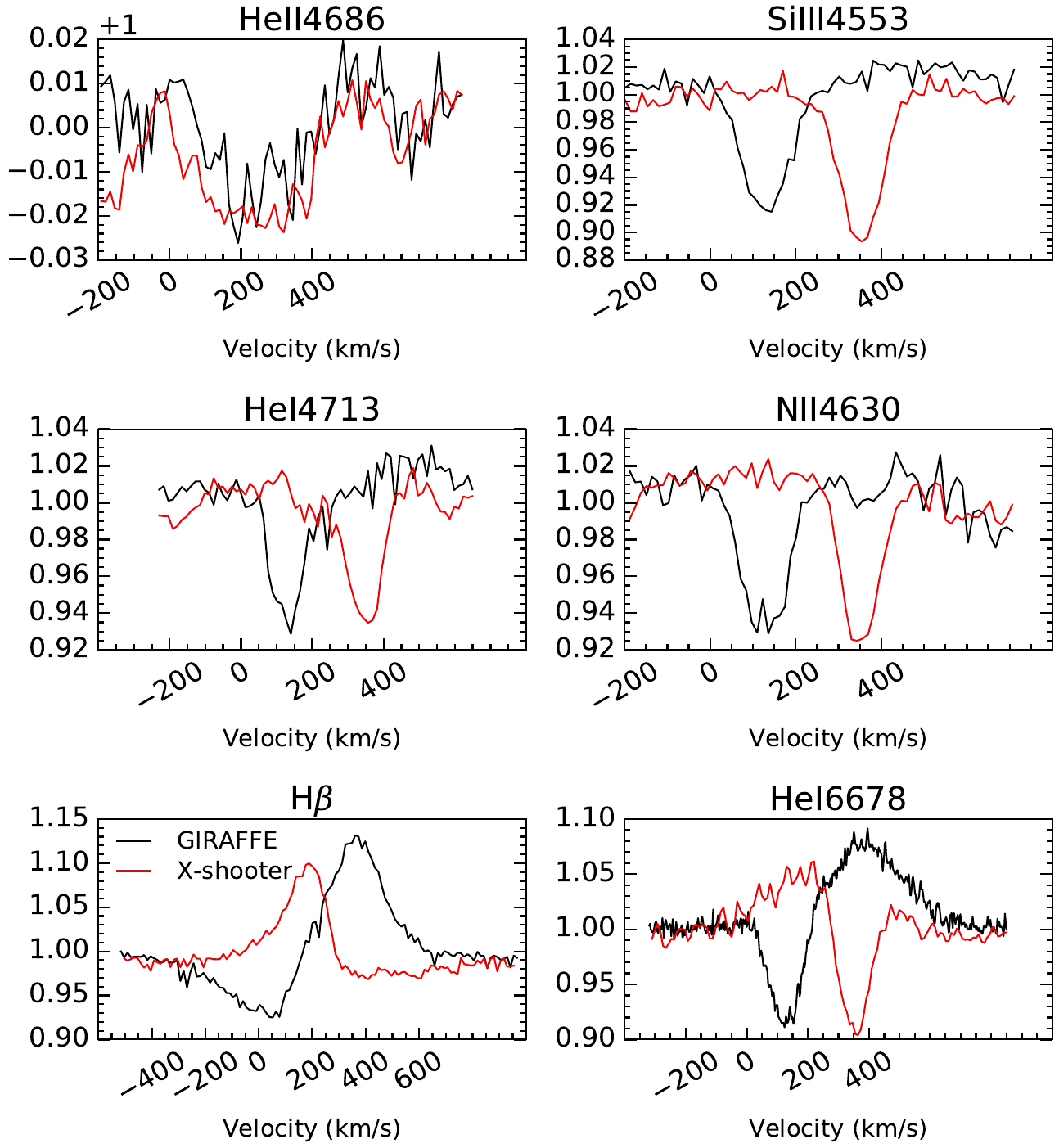}
    \caption{Radial velocity shift between Flames/GIRAFFE and X-shooter spectra of \mbox{Sk\,-71$^{\circ}$\,35}}
    \label{fig:RV7135}
\end{figure}

We checked for RV variations in the spectra by comparing the X-shooter and GIRAFFE spectra as shown in Fig.\,\ref{fig:RV7135}.
The narrow absorption lines show an RV variation of $\approx215$ km\,s$^{-1}$ between the two epochs. The UV spectra also exhibit RV shifts relative to the optical spectra.
For the primary star, the maximum $ \delta_\text{RV}$ from the available observations is $\approx290$ km\,s$^{-1}$. On the other hand, the RV variations in the broad \ion{He}{ii} lines are much smaller. The maximum $\Delta_\text{RV}$ for the secondary from the UV and optical spectra is  $<70$ km\,s$^{-1}$, indicating that the partially stripped star has a much lower mass than the Oe star.  

The disk emissions in H$\alpha$ and H$\beta$ are found to be moving in the opposite direction to the narrow lines, confirming their secondary origin.
Moreover, asymmetric double peak-like emissions are visible in \ion{He}{i} lines in the red part of the optical spectra. For example, \ion{He}{i\,$\lambda$6678} shows both narrow absorption (from the stripped primary) and disk emission component (from the Oe star) moving in opposite directions (see Fig.\,\ref{fig:RV7135}). 
This kind of emission plus absorption features from both components are also imprinted in the Paschen lines.
In all these cases, the RV shifts between two stars are in agreement with the measure from the \ion{He}{i} absorption lines in the blue part of the optical spectra where we can see both narrow and broad components.

Only three epochs of spectra are available for \mbox{Sk\,-71$^{\circ}$\,35}. The details of the RV measurements are given in Sect.\,\ref{appendix:RV2}. However, since it is a bright star, we looked at the TESS light curves to determine the orbital period.  The light curve exhibits ellipsoidal modulation, suggesting that it is likely caused by the gravitational distortion of one of the two stars--likely the partially stripped primary star, which is puffier and more prone to distortion. We derived a binary orbital period of $\approx9.4$ days for the system. Details of the light curve analysis can be found in  Sect.\,\ref{appendix:tess}. Using the period derived from the TESS light curve and using 3  RV measurements we carried out the orbital analysis of the system using  \texttt{rvfit} code (see Sect.\,\ref{appendix:rvfit}).


We performed the spectral analysis of \mbox{Sk\,-71$^{\circ}$\,35} by following a similar method described in \cite{Ramachandran2023}. The main diagnostic to constrain the temperature of the primary in \mbox{Sk\,-71$^{\circ}$\,35} is the He and Si ionization balance based on \ion{He}{i}\,/\,\ion{He}{ii} and \ion{Si}{iii}\,/\,\ion{Si}{iv} line ratios. The secondary star is much hotter than the primary in this system, as indicated by the broad \ion{He}{ii} lines in the optical and \ion{N}{V} lines in the UV. From the broad components of the \ion{He}{i} and \ion{He}{ii} lines, we constrained the temperature of the Oe star. For the surface gravity, the pressure-broadened wings of the Balmer lines are the primary diagnostics. We considered H$\gamma$ and H$\delta$ since they are less impacted by wind and disk emissions. Nevertheless, both H$\gamma$ and H$\delta$ contain primary and secondary contributions. As a result, the luminosity ratios and surface gravities in the primary and secondary models were concurrently modified to match with the observations. Since the ionization balance also reacts to gravity, we simultaneously re-adjusted $T_\ast$ and $\log\,g_\ast$ to achieve a good fit to the observed spectra.

\begin{figure}  
\vspace{0.5cm}
\includegraphics[scale=0.8]{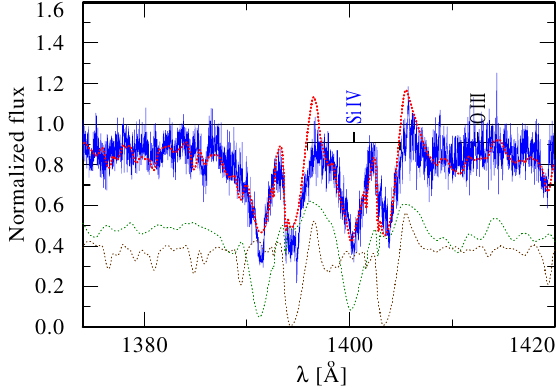}
\caption{\ion{Si}{iv} P\,Cygni line of \mbox{Sk\,-71$^{\circ}$\,35}. The composite
model (dashed red) is the weighted sum of the stripped star primary (dotted brown)
and Oe secondary (dashed green) model spectra with an RV shift.}
\label{fig:SiIV}
\end{figure}

Unlike the other two systems discussed before, the secondary component in \mbox{Sk\,-71$^{\circ}$\,35} has a stronger contribution to the UV spectrum. 
Moreover, we can clearly distinguish the wind components from both primary and secondary due to the temperature and RV differences.
The P\,Cygni lines of \ion{C}{iv\,$\lambda\lambda$1548--1551} and \ion{N} {v\,$\lambda\lambda$1238--1242} are coming from the hotter secondary. 
On the other hand, \ion{Si}{iv\,$\lambda\lambda$1393--1403} P\,Cygni profiles are visible from both components and well-separated with an RV difference of $\approx 200$ km\,s$^{-1}$ (see Fig.\,\ref{fig:SiIV}). We measure $\varv_\infty$ from the blue edge of these lines for both the primary and secondary. For the secondary, this measurement of $\varv_\infty$ is consistent with the value from the \ion{N}{v} and \ion{C}{iv} profiles. The mass-loss rate ($\dot{M}$) is estimated for both components from the strength of the P\,Cygni profiles. For the secondary, we have to incorporate an additional X-ray field into the model with an X-ray luminosity of $\log L_{\rm X} =32.8$\,erg\,s$^{-1}$ to reproduce \ion{N}{v}. This is in agreement with the upper limit on X-ray luminosity of \mbox{Sk\,-71$^{\circ}$\,35}, $\log{L_X}<34$ in 0.2-12.0 keV

We constrained the CNO abundances for the primary and secondary components in the system. For the primary, we use CNO lines in the optical spectra. 
 The N abundance is determined by analyzing multiple \ion{N}{ii} and \ion{N}{iii} lines. Although we found an enrichment by a factor of 16 compared to the LMC baseline value, the enrichment is actually much lower than in the other two newly found stripped stars. The C and O abundances are estimated from several \ion{C}{iii} and \ion{O}{ii} lines in the optical spectra. The CNO abundances for the secondary are constrained from the UV lines (\ion{C}{iii}, \ion{N}{iii}, \ion{N}{iv}, and \ion{O}{iv}).
 In addition to the CNO abundances, we varied the H mass fraction ($X_{\rm H}$) in the primary models between $0.5$ and $0.73$. Our best-fit model to the observations suggests $X_{\rm H}=0.7$, does not favor any signs of surface-He enrichment or H depletion.  
 
\begin{table}
	\caption{Fundamental parameters and abundances derived for  \mbox{Sk\,-71$^{\circ}$\,35}}
	\label{table:parameters3}
	\centering
	\renewcommand{\arraystretch}{1.6}
	\begin{tabular}{lcc}
		\hline 
		\hline
		\vspace{0.1cm}
&	Stripped star                                     &     Oe star           \\
		\hline 
    Spectral type & B1 Ia  & O9 Ve\\
		$T_{\ast}$ (kK)                                & $21^{+2}_{-2}$       &  $32^{+3}_{-3}$ \\
 $T_{2/3}$ (kK)                                 & $20^{+2}_{-2}$ &  $31.8^{+3}_{-3}$ \\
   
		$\log g_\ast$ (cm\,s$^{-2}$)    & $2.6^{+0.2}_{-0.2}$   &   $3.8^{+0.2}_{-0.2}$   \\
 $\log g_{2/3}$ (cm\,s$^{-2}$)    & $2.51^{+0.1}_{-0.2}$    &    $3.79^{+0.2}_{-0.2}$   \\
  $\log g_\mathrm{true}$\tablefootmark{\dag} (cm\,s$^{-2}$)    & $2.65^{+0.1}_{-0.2}$   &   $3.85^{+0.2}_{-0.2}$   \\
		$\log L$ ($L_\odot$)                           & $4.93^{+0.1}_{-0.1}$  &   $5.12^{+0.2}_{-0.2}$    \\ 
		$R_\ast$ ($R_\odot$)                           & $22.1^{+2}_{-2}$    &    $11.8^{+2}_{-2}$      \\
  $R_{2/3}$  ($R_\odot$)                               & $24.3^{+2}_{-2}$  &   $11.9^{+2}_{-2}$    \\
		$\log \dot{M}$ ($M_\odot \mathrm{yr}^{-1}$)    & $-7.0^{+0.2}_{-0.2}$  &$-7.1^{+0.2}_{-0.2}$  \\
		$\varv_{\infty}$ (km\,s$^{-1}$)     & $400^{+50}_{-50}$ & $900^{+100}_{-100}$  \\ 
  $\beta$                               &   0.8          & 0.8  (fixed)      \\
		$D$                               &   10          & $10$  (fixed)      \\
		$\varv \sin i$ (km\,s$^{-1}$)       & $85^{+10}_{-10}$   &  $250^{+80}_{-80}$      \\
		$\varv_{\mathrm{mac}}$ (km\,s$^{-1}$)         & $30^{+10}_{-10}$   &    50 (fixed)   \\
  $\varv / \varv_{\mathrm {crit}}$       & $\gtrsim 0.4$   &  $\gtrsim 0.4$      \\
		$\xi$ (km\,s$^{-1}$)            & $15$     &      $15$  (fixed)   \\
		$X_{\rm H}$ (mass fr.)                         & $0.7^{+0.04}_{-0.1}$  &   0.737 (fixed)   \\
		$X_{\rm He}$ (mass fr.)                         & $0.29^{+0.1}_{-0.1}$  &   0.26 (fixed)   \\
		$X_{\rm C}/10^{-5}$ (mass fr.)                 & $10^{+2}_{-2}$       &  $3^{+1}_{-1}$    \\
		$X_{\rm N}/10^{-5}$ (mass fr.)                 & $100^{+10}_{-10}$  & $20^{+7}_{-7}$        \\
		$X_{\rm O}/10^{-5}$ (mass fr.)                 & $64^{+10}_{-10}$      &   $164^{+100}_{-50}$    \\
		$E_{\mathrm{B-V}}$ (mag)                                & $0.14^{+0.02}_{-0.02}$ &    \\
		$M_\mathrm{spec}$ ($M_\odot$)                  & $7.8^{+3.6}_{-3}$    &  $35.7^{+18}_{-14}$      \\
		$\log\,Q_{\mathrm H}$ (s$^{-1}$)   &47.2  & 48.4   \\
		$\log\,Q_{\mathrm {He\,\textsc{ii}}}$ (s$^{-1}$)   & - &  42   \\
  $P_{\mathrm{orb}}$ (days)     & 9.398 \\
$R_\mathrm{RL}\,(R_\odot)$ & $17.0^{+9}_{-7}$ & $33.8^{+19}_{-15}$\\
$f_\mathrm{RL}$ & $1.29^{+0.71}_{-0.56}$ & $0.35^{+0.2}_{-0.15}$\\
		\hline
	\end{tabular}
 \tablefoot{ \tablefoottext{$\dag$} {$\log (g_\ast +(\varv \sin i)^{2} /R_\ast)$}
 
 }
\vspace{-0.2cm}
\end{table}

 In  \mbox{Sk\,-71$^{\circ}$\,35}, the partially stripped star contributes approximately 65\% of the light in the optical, but only 30\% in the UV. We adjusted the luminosities of primary and secondary models such that the composite model SED matches the photometry and flux-calibrated UV spectra (Fig.\,\ref{fig:sedfit3}). Here the model flux is diluted with
the adopted LMC distance modulus of 18.48 mag \citep{Pietrzynski2019}.
The final best-fit model and SED for 2dFS\,2553 are shown in Figs.\,\ref{fig:optfit3}, \ref{fig:uvfit3}, and \ref{fig:sedfit3} respectively.
The stellar and wind parameters as well as surface abundances derived for both the stripped star and the Oe star are given in Table\,\ref{table:parameters3}.
The system  \mbox{Sk\,-71$^{\circ}$\,35} is found to be semi-detached where the less massive partially stripped star is filling its Roche lobe and transferring mass.

\section{Discussions}
\label{discussion}

\subsection{Stellar masses}\label{sec:discussion-mass}
\begin{figure}
    \centering
    \includegraphics[width=0.85\linewidth]{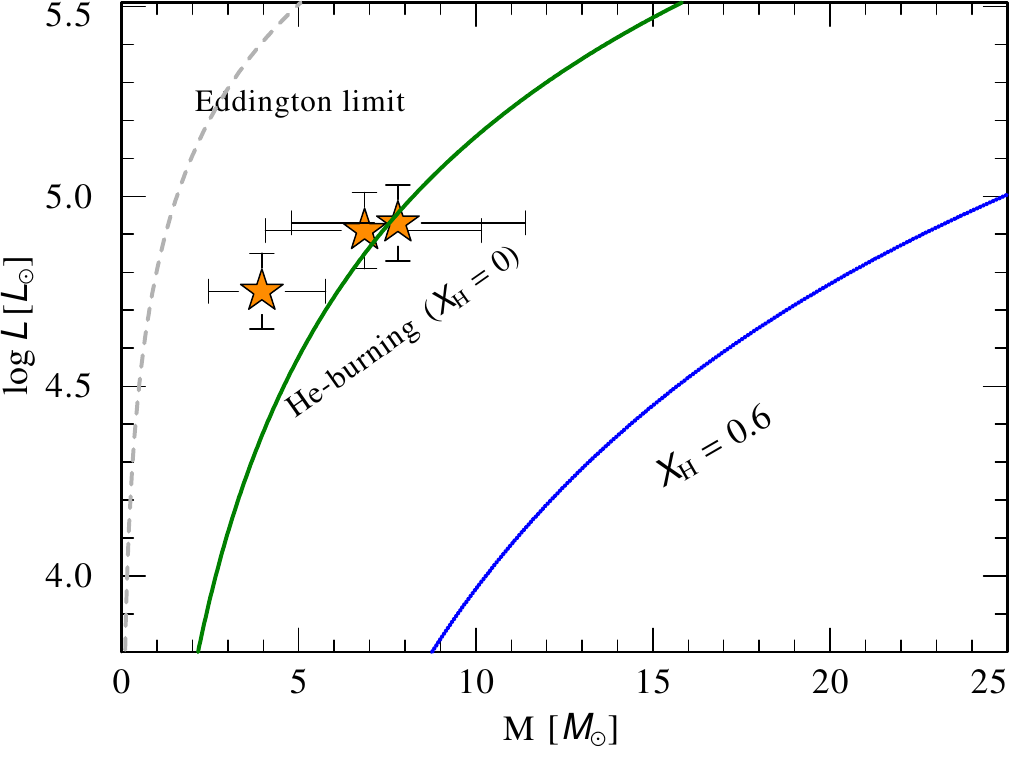}
    \caption{Luminosity versus stellar mass for our newly discovered stripped stars. For comparison, theoretical relations for chemically homogeneous stars with $60$\% and zero surface hydrogen as well as the Eddington limit are shown as well \citep{Graefener2011}.}
    \label{fig:ML}
\end{figure}

We determined the spectroscopic mass from gravity and stellar radius using the relation $g_\mathrm{true}=G\,M_\ast\,R_\ast^{-2}$ for each of the components in our sample binaries. Here $g_\mathrm{true}$ is determined via the measured $\log\,g_\ast$ from the analysis, after accounting for centrifugal acceleration using the relation $\log (g_\ast +(\varv \sin i)^{2} /R_\ast)$\citep{Repolust2004}.   
The mass of the primary in 2dFS\,163 is determined to be only half of the mass of the secondary, which contributes significantly less to the observed spectra compared to the primary.
 The mass of the primary is estimated to be $\sim$$4\,M_{\odot}$, making it a candidate for an intermediate-mass stripped star.
The spectroscopic masses of the stripped stars in 2dFS\,2553 and \mbox{Sk\,-71$^{\circ}$\,35} are relatively high, approximately $7\,M_{\odot}$.
Both 2dFS\,163 and 2dFS\,2553 exhibit spectral characteristics similar to early-type O stars which usually have masses greater than 30\,$M_{\odot}$ \citep[for e.g.,][]{Martins2005}. However, the estimated masses of these primaries are considerably smaller, suggesting He-core burning and substantial removal of their outer layers. In addition, their smaller radii ($R< 6\,R_{\odot}$) suggest that they might be in an ongoing contracting phase. Similarly, the mass of the stripped star in \mbox{Sk\,-71$^{\circ}$\,35} is way lower than expected for an early B supergiant. In this case, the stripped star has a much larger radius and short orbital period, with a Roche lobe filling factor $>1$,  indicating an ongoing mass transfer phase. 
This is in agreement with the observed ellipsoidal modulations in the light curve.

The secondary Be star in 2dFS\,2553 is twice as massive as the stripped star.
For 2dFS\,2553 the mass ratio determined through orbital analysis using PHOEBE (see Sect.\,\ref{appendix:PHOEBE}) is lower, but it falls within the range of error and is consistent. 
In \mbox{Sk\,-71$^{\circ}$\,35}, the Oe star is found to be 4.6 times more massive than the stripped star, and the orbital analysis is also in agreement with a high mass-ratio (see Sect.\,\ref{appendix:rvfit}). \update{ It should be noted that since the lines of secondaries are weaker in the observed spectra, the true uncertainty in the estimated masses of these objects may be higher.}
For stripped primaries, we found agreement between spectroscopic masses, and those obtained from evolutionary models (see Sect.\,\ref{sec:discussion-evo}) and orbital analysis (Sections\,\ref{appendix:PHOEBE}, \ref{appendix:rvfit}).

The positions of the primaries in the mass-luminosity diagram shown in Fig.\,\ref{fig:ML} are also in agreement with envelope stripping.
All three stripped stars are over-luminous for their current masses and hence cluster near the curve corresponding to fully stripped, He-burning stars.
A similar M-L position is also reported for SMCSGS-FS\,69 \citep{Ramachandran2023}. All of the observed characteristics indicate that our discovered systems are post-interaction binaries, wherein the primary star has undergone significant mass transfer and is burning He in its core, while the secondary star has experienced both mass and angular momentum gain.
However, contrary to the usually assumed picture, these stars
are not fully stripped, possess a considerable H-rich envelope, and have quite high
 luminosity and low gravity, hence effectively hiding among the standard OB-type population.
 
\subsection{Chemical abundances}
\begin{figure}
    \centering
    \includegraphics[width=0.99\linewidth]{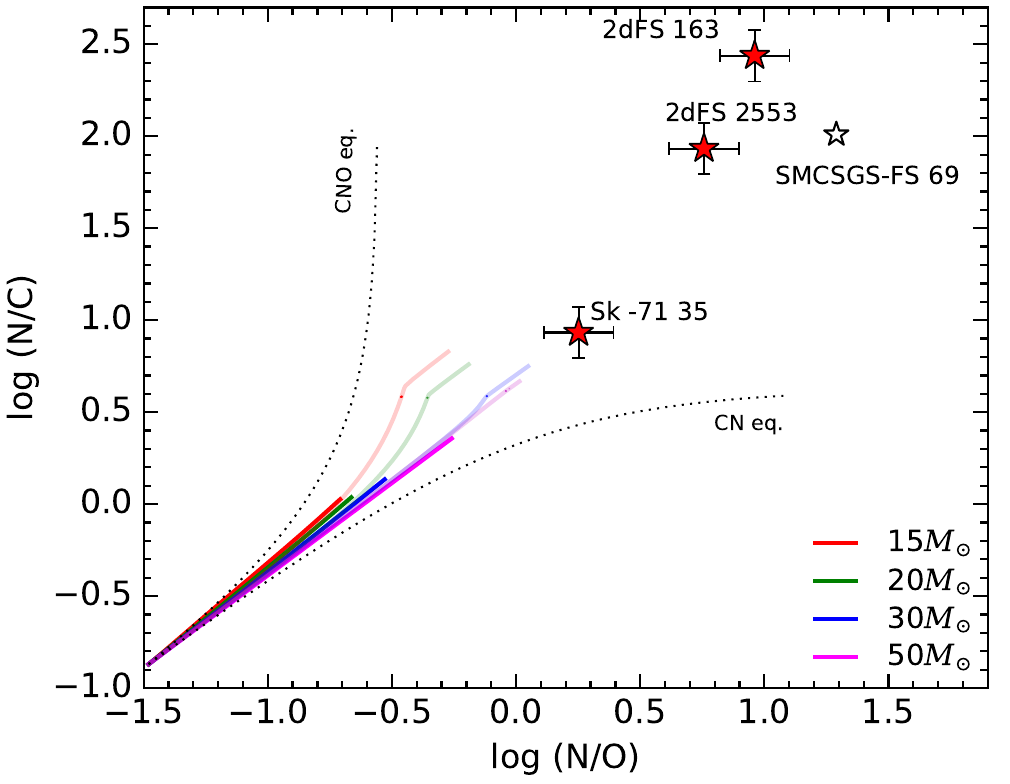}
    \caption{ log (N/C) vs. log (N/O) abundances (by number) for the stripped primaries in our sample. For comparison the location of the previously reported partially stripped star in the SMC, SMCSGS-FS\,69 \citep{Ramachandran2023} is marked.
    The dotted lines indicate the expected trends for the case of partial CN and complete CNO equilibrium. Solid lines are based on evolutionary tracks from  \cite{Brott2011} for different initial masses and initial rotational velocities corresponding to  0 and 300\,km/s.}
     \label{fig:CNO}
\end{figure}
All three partially stripped stars show signs of nitrogen enrichment and CO depletion. Noticeably, in the case of two compact stripped stars found in the SMC, the [N/C] and [N/O] ratios are significantly high as shown in Fig.\,\ref{fig:CNO}. The CNO abundances are also found to be in alignment with those reported for the previously reported partially stripped star in the SMC, SMCSGS-FS\,69 \citep{Ramachandran2023}.
On the other hand, the cooler stripped star in the LMC, \mbox{Sk\,-71$^{\circ}$\,35}, shows only a rather mild N enrichment and CO depletion. This could be because the system may be undergoing mass transfer. 
In any case, these CNO abundances are very different than those expected for OB stars based on evolutionary models \citep[e.g.,][]{Brott2011} as well as compared to observations of OB stars in the Magellanic Clouds \citep{Hunter2009}. Even assuming a large initial rotation, which can result in chemical mixing and nitrogen enrichment, is not enough to explain the observed CNO ratios in our sample objects.

The high nitrogen-to-carbon and nitrogen-to-oxygen ratios observed in our sample stars instead suggest that the observed significant chemical evolution is rooted in envelope stripping. Furthermore, strong surface He enrichment is observed in 2dFS\,163, indicating that it has undergone intense envelope stripping, exposing the deeper layers.

\update{It is also worth noting that the total CNO abundances of two SMC stars are higher than the baseline values adopted from \cite{Hunter2007,Trundle2007}. For 2dFS\,163, $\Sigma$CNO (mass fraction) =$1.8\times10^{-3}$ and for 2dFS\,2553 this is $2.4\times10^{-3}$, whereas the baseline value is $\approx 1.4\times10^{-3}$.} This is possibly due to a higher scatter in the initial abundances of stars across SMC from the baseline value. Within the XshootU collaboration, \citet[XShootU V]{Martins2024} studied the CNO abundance of O stars and found a similar higher total CNO abundance from the baseline values for some of the SMC stars.  Furthermore, \cite{Ramachandran2021} have also reported a significant dispersion in CNO abundances among O stars in the SMC Bridge region.

\subsection{Wind parameters of stripped stars}
\begin{figure}
    \centering
    \includegraphics[width=0.99\linewidth]{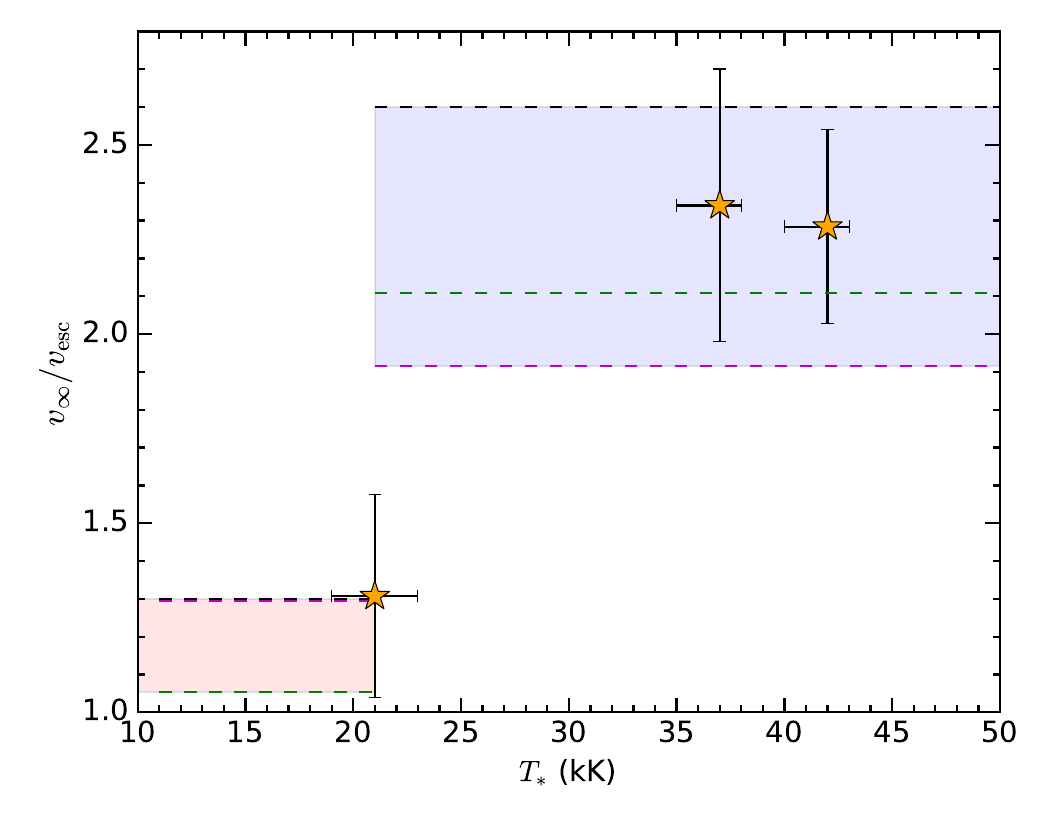}
    \caption{Terminal velocity to escape velocity ratio for our sample stripped stars: For comparison, we also plotted the $\varv_\infty / \varv_{\mathrm{esc},\Gamma}$ ratios observed for OB stars. 
        The black dashed lines represent $\varv_\infty / \varv_{\mathrm{esc},\Gamma}$  for Galactic OB stars as reported in the study by \cite{Lamers1995}. The ratios are different for stars with $T_{\ast}\gtrsim21\,$kK (blue region)  for stars cooler than 21\,kK (red region).
        The green dashed lines represent the same ratio but scaled to SMC metallicity, as described in \cite{Leitherer1992}, whereas the magenta dashed lines indicate a steeper metallicity scaling found by \cite{VinkSander2021}. }
    \label{fig:vinfvesc}
\end{figure}

\begin{figure}
    \centering
    \includegraphics[width=0.99\linewidth]{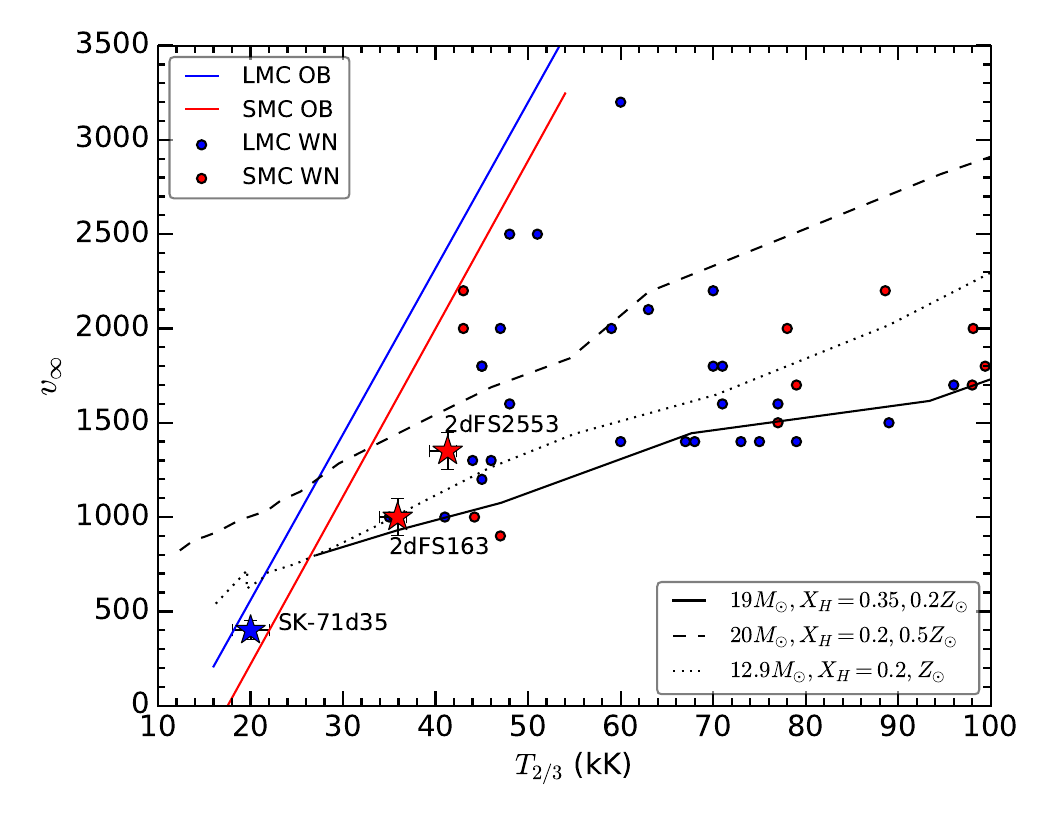}
    \caption{Terminal velocity vs. Temperature for our sample stripped stars. The position of WR stars in the Magellanic clouds \citep{Shenar2016,shenar_2019_wolfrayet}, theoretical WN model sequences (black lines) from \citet{Sander2023}, and the correlation for OB stars from the ULLYSES sample \citep{Hawcroft2023} are shown for comparison.}
    \label{fig:vinfT}
\end{figure}

\begin{figure*}
    \centering
    \includegraphics[width=0.48\linewidth,trim={0cm 0cm 0cm 0cm},clip]{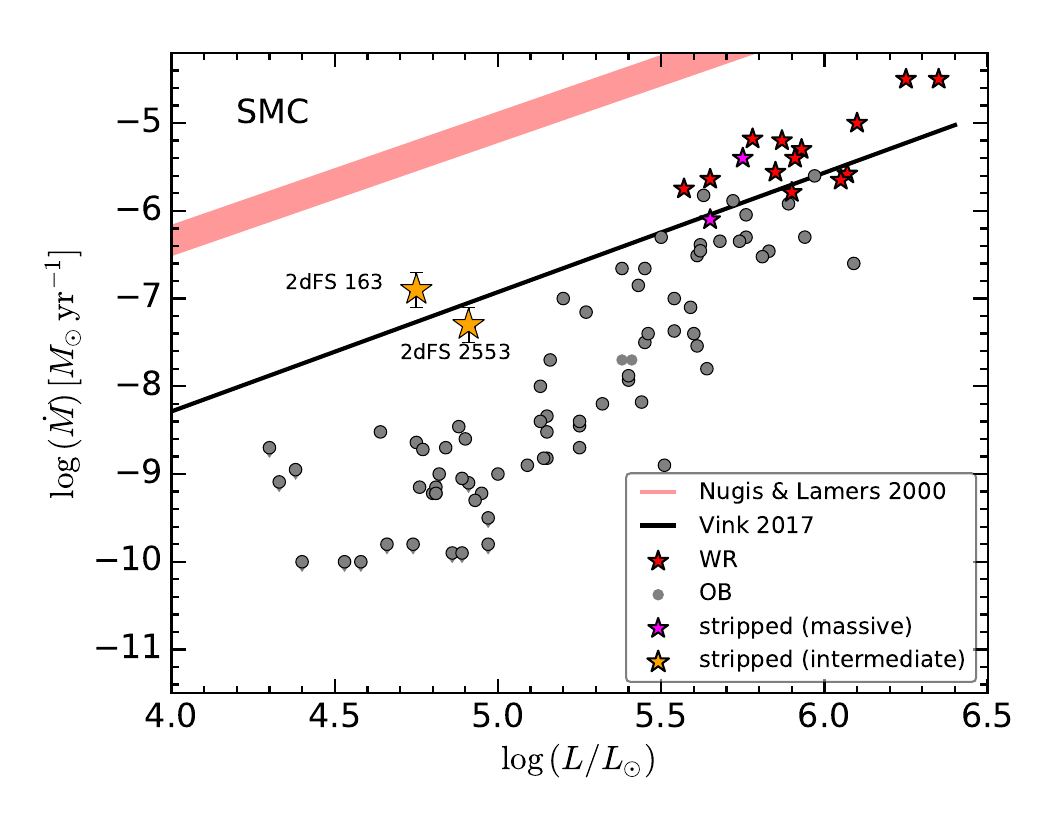}
    \includegraphics[width=0.48\linewidth,trim={0cm 0cm 0cm 0cm},clip]{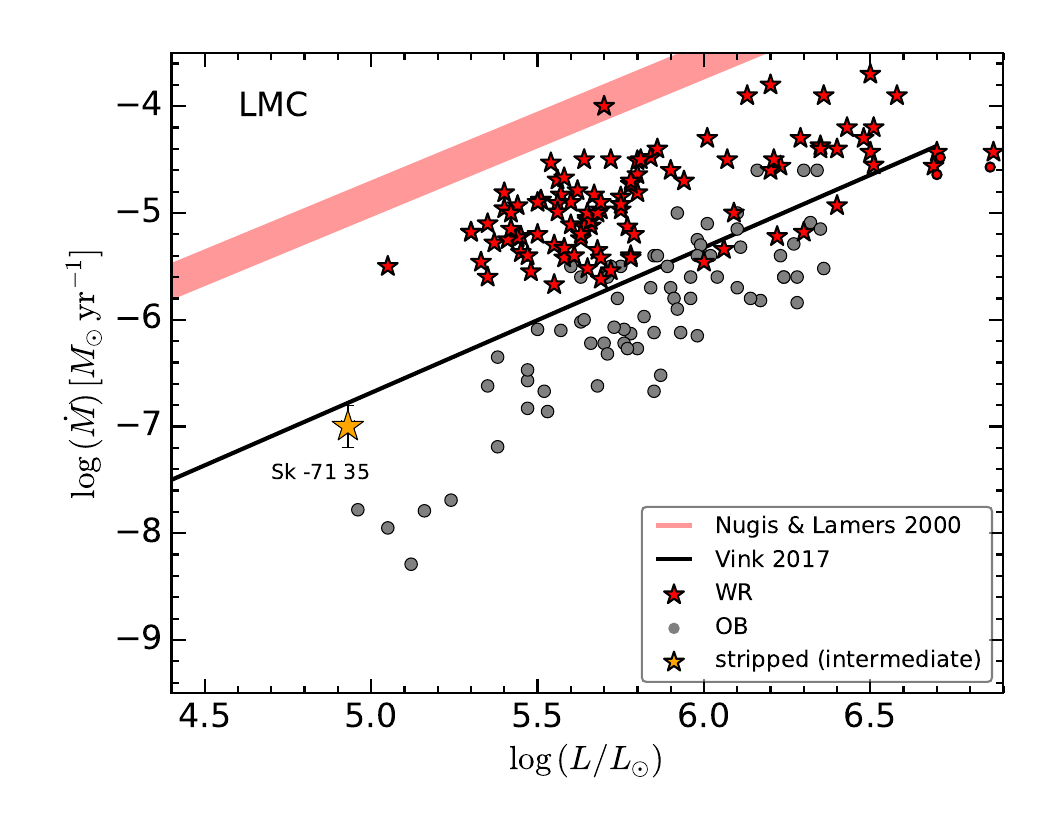}
    \caption{Mass-loss rates vs. luminosity relation for stripped stars in the SMC (left) and LMC (right). Stripped stars in the intermediate mass range discovered in this study are shown as orange stars. Massive stripped stars in binaries from \cite{Pauli2022b} and \cite{Rickard2023} are shown as magenta stars.  For comparison, the locations of OB stars \citep{Bouret2013,Ramachandran2019,Bouret2021, Rickard2022,Brands2022} and WRs \citep{Hainich2014,Hainich2015,Shenar2016,shenar_2019_wolfrayet,Bestenlehner2014} are plotted from the literature. The black solid line shows the mass-loss rate prescriptions for stripped helium stars by \cite{Vink2017} and the red line corresponds to the mass-loss recipe for WR stars based on \cite{Nugis2000}. 
    }
    \label{fig:mdot}
\end{figure*}

\subsubsection{Terminal velocity}\label{sec:discussion-vinf}
Given the absence of prior measurements on the terminal wind velocities of partially stripped stars, it was a priori not clear whether they follow common correlations for OB stars with respect to the effective escape velocity 
\begin{equation}
  \varv_{\mathrm{esc},\Gamma} = \sqrt{\frac{2 G M}{R} \left(1 - \Gamma_\text{e}\right)}
\end{equation}
or the stellar temperature. In Fig.\,\ref{fig:vinfvesc}, we show the $\varv_\infty$ to $\varv_{\mathrm{esc},\Gamma}$ ratio for each of the stripped stars in our sample at their corresponding temperature. The relationship between $\varv_\infty$ and $\varv_{\mathrm{esc},\Gamma}$ has been investigated in both theoretical and observational studies for massive stars in our Milky Way and is found to be $\varv_\infty$/$\varv_{\mathrm{esc},\Gamma}\simeq$\,2.6 for stars with $T_\ast\geq$\,21\,kK, while for stars with $T_\ast <$\,21\,kK the ratio is $\approx$1.3 \citep{Lamers1995,Kudritzki2000}. The terminal velocity is also expected to scale with the metallicity, $\varv_\infty \, \propto (Z/Z_{\odot})^{\alpha}$, where $\alpha=0.13$ \citep{Leitherer1992}. Theoretical calculations by \citet{VinkSander2021} based on Monte Carlo models employing the \citet{MuellerVink2008} approach find a difference in the $\varv_\infty(Z)$-scaling for the two temperature regimes with $\alpha=0.19$ for the hot side -- in line with the empirical findings by \citet{Hawcroft2023} -- and an almost vanishing dependency of $\alpha=0.003$ for the cooler effective temperatures. As evident from Fig.\,\ref{fig:vinfvesc}, the hot stripped stars in our study align well with the predictions. 
\citet{Hawcroft2023} studied the $\varv_\infty$ to $\varv_{\mathrm{esc},\Gamma}$ ratio for the OB stars in the ULLYSES sample and found a significant scatter for both LMC and SMC stars. However, while using their newly determined values for $\varv_\infty$, the values for $\varv_{\mathrm{esc},\Gamma}$ was based on a diverse set of literature sources. We thus expect some of the values to change with the availability of the ULLYSES and XShootU data which could give better-quality spectra or reveal previously unknown binaries. Within the XShootU collaboration, Bernini-Peron et al. (2024, in prep, XShootU XI?) recently studied a sample of B supergiants in the SMC, yielding a considerably lower scatter for $\varv_\infty$ to $\varv_{\mathrm{esc},\Gamma}$ and only a few, distinct outliers. Moreover, their findings also point towards a rather weak scaling of $\varv_\infty$ with $Z$, even above $21\,$kK, at least for supergiants. The results for our stripped stars, which are more supergiant-like in their surface gravities and winds, seem to align with these findings. 

Although the $\varv_\infty$ to $\varv_{\mathrm{esc},\Gamma}$ ratio of stripped stars is in agreement with OB stars, in particular, the $\varv_\infty$ for the two hot stripped stars in the SMC is much lower than observed for OB stars of similar temperatures as shown in Fig.\,\ref{fig:vinfT}. For comparison, we also plot the determined parameters for WN stars in the Magellanic Clouds \citep{Hainich2014,Hainich2015,Shenar2016,shenar_2019_wolfrayet} and a few theoretical WN model sequences from \citet{Sander2023} including a previously unpublished dataset with SMC-like metallicity ($Z = 0.2\,Z_\odot$) and $X_\text{H} = 0.35$. The optically thick WR winds show a flatter slope with $L/M$-ratio, metallicity, and hydrogen abundance mainly producing an overall shift. The position of the two hot stripped stars in the $\varv_\infty$-$T_{2/3}$ plane is clearly shifted towards the locations of the cooler WN stars, but since their winds are not optically thick (cf.\ discussion below) we would not expect them to follow the scalings for the classical WRs.

\subsubsection{Mass-loss rates}\label{sec:discussion-mdot}

Multiple studies have determined empirical mass-loss rates for WR and OB stars in the Galaxy and the Magellanic Clouds. Yet, the wind mass-loss rates of intermediate-mass stripped stars have remained unknown for the following reasons: i) Only a small number of potential candidates have been identified so far based on observations.  ii) Most of these detections are based on optical spectroscopy, and no UV spectra were available to measure mass-loss rates iii) The winds of these stars are not as powerful as those of WR stars, so typically there are no wind emission lines visible in the optical spectra.  iv) Another difficulty in accurately measuring wind mass-loss rates in intermediate-mass stripped stars from optical spectra is the presence of companions with disks, which can contaminate H$\alpha$.

In this work, we report the first mass-loss rate estimates for intermediate-mass stripped stars from a combined UV and optical spectral analysis. The derived mass-loss rates are compared with those of WR and OB stars in Fig.\,\ref{fig:mdot}, showing the $\dot{M}$-$L$ plane for both the LMC and SMC. The stripped star mass-loss rates found in this study are on the order of $10^{-7}\,M_\odot \mathrm{yr}^{-1}$, considerably higher than the mass-rates observed for ``normal'' OB stars with a similar luminosity. This is also significantly higher than the reported upper limits for stripped star mass-loss rates in \cite{Gotberg2023} based on the optical \ion{He}{ii$\lambda4686$} absorption line. 

To put their spectral appearance in context to the more massive WR stars, \update{it is helpful to consider quantities that describe the relative strengths of their line emission such as their transformed radii $R_\text{t} 
  = R_\ast  ( \varv_\infty /2500\,\mathrm{km s^{-1}} )^{2/3} (\dot{M} \sqrt{D} / 10^{-4} M_\odot \mathrm{yr}^{-1})^{-2/3}$\citep{Schmutz1989} or their transformed mass-loss rate  $
  \dot{M}_\text{t} = \dot{M} \sqrt{D} \cdot ( 1000\,\mathrm{km s^{-1}} / \varv_\infty ) (  10^6 L_\odot/L)^{3/4}$\citep{Graefener2013}. }
Since $R_\text{t}$ was historically derived from observed scalings between different models, larger values actually mean less emission. For our targets, we obtain values of $\log R_\text{t} = 1.97$ (2dFS\,163), $2.44$ (2dFS\,2553), and $2.35$ (\mbox{Sk\,-71$^{\circ}$\,35}). All of these values are above the limit for which emission-line spectra are expected \citep[see, e.g.,][]{Shenar2020}, although the coolest stripped star \mbox{Sk\,-71$^{\circ}$\,35} shows a P\,Cygni profile in H$\alpha$ according to our models. This is also reflected in the slightly more intuitively $\dot{M}_\text{t}$, which describes the mass-loss rate the star would have if it had a smooth wind, a terminal velocity of $1000\,\mathrm{km}\,\mathrm{s}^{-1}$ and a luminosity of $10^{6}\,L_\odot$. Indeed, \mbox{Sk\,-71$^{\circ}$\,35} has the highest value in $\log \dot{M}_\text{t}$, namely $-5.1$ (compared to $-5.3$ for 2dFS\,163 and $-6.1$ for 2dFS\,2553), indicating the most dense wind of the sample. Still, this is too low to reach the regime of optically thick winds, which starts above $\log \dot{M}_\text{t} \approx -4.5$ for classical WR stars \citep{Sander2023}.

\begin{figure}
    \centering
    \includegraphics[width=0.95\linewidth]{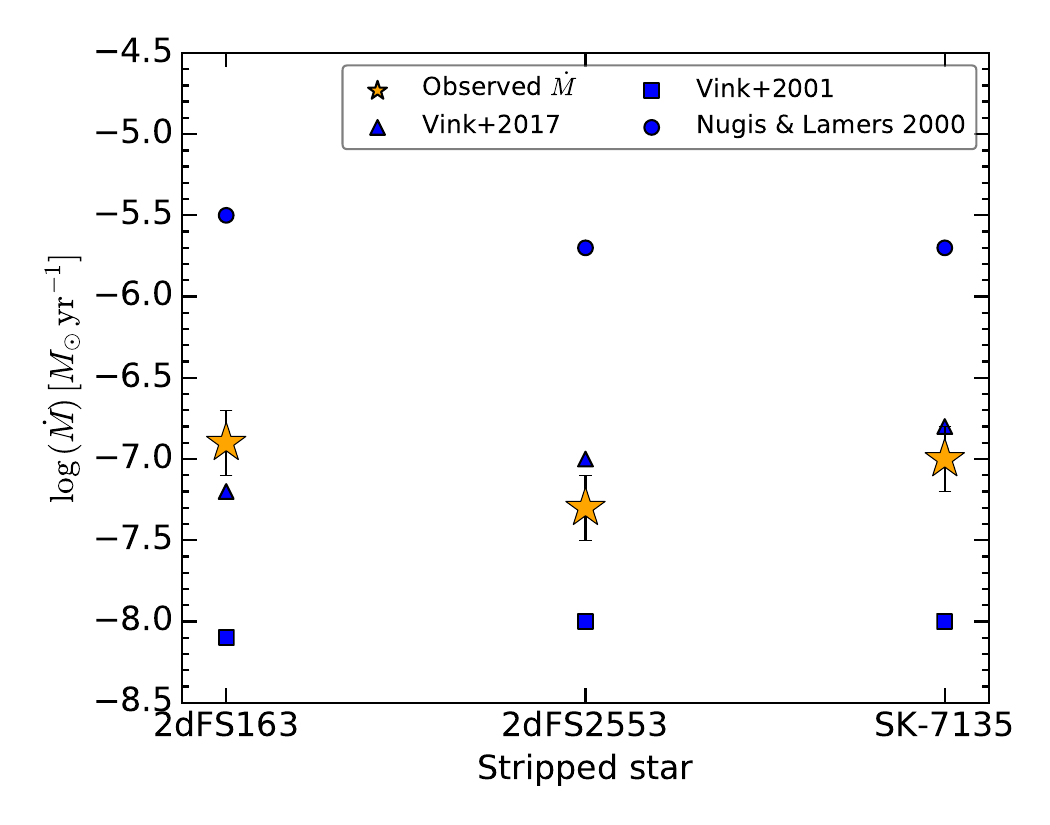}
    \caption{Estimated mass-loss rates for stripped stars from the spectral analysis compared to commonly used recipes.  }
    \label{fig:mdot_comp}
\end{figure}

All of our derived mass mass-loss rates are roughly within factors of $2$-$3$, in agreement with the \citet{Vink2017} predictions for stripped stars, despite some objects being considerably cooler than the assumed fixed temperature of $50\,$kK in their models. The predictions from \citet{Vink2017} yield intermediate values between the empirical WR and OB mass-loss rates in both LMC and SMC. Based on the derived parameters for our three targets, we applied typically used mass-loss recipes such as \cite{Nugis2000} for WR stars, and \cite{Vink2000,Vink2001} for OB stars in addition to \cite{Vink2017} predictions for stripped stars. These values are then compared to the observed mass-loss rates as shown in Fig.\,\ref{fig:mdot_comp}.
In all cases, \cite{Vink2000,Vink2001} mass-loss rates underestimate the observed values by roughly an order of magnitude, while \cite{Nugis2000} overestimate them by more than a factor of 30. The derived parameters, in particular the optically thin winds and the comparably low temperatures, yield that cWR-wind recipes, including \citet{SanderVink2020}, are inapplicable for all of our sample stars.
On the other hand, during stripped phases, WR or OB mass-loss prescriptions are frequently employed in evolutionary models and population synthesis. 
Their significant disparities with the observed mass-loss rates will have a substantial impact on the final evolutionary fate and the mass of the compact object. For e.g, \cite{Gilkis2019} showed that the mass-loss prescriptions for stripped stars assumed in the evolutionary models influence the fraction of stripped
supernovae of types IIb and Ibc. Including accurate measurements of wind mass-loss rates of intermediate-mass stripped stars into evolutionary models and population synthesis studies could provide valuable insights into stellar evolution processes at this range of masses. 

\subsection{Evolutionary nature and implications of the discovered systems}
\label{sec:discussion-evo}

\begin{figure*}
        \centering
        \includegraphics[trim={2.cm 0cm 5.5cm 0cm},clip,width=0.95\textwidth]{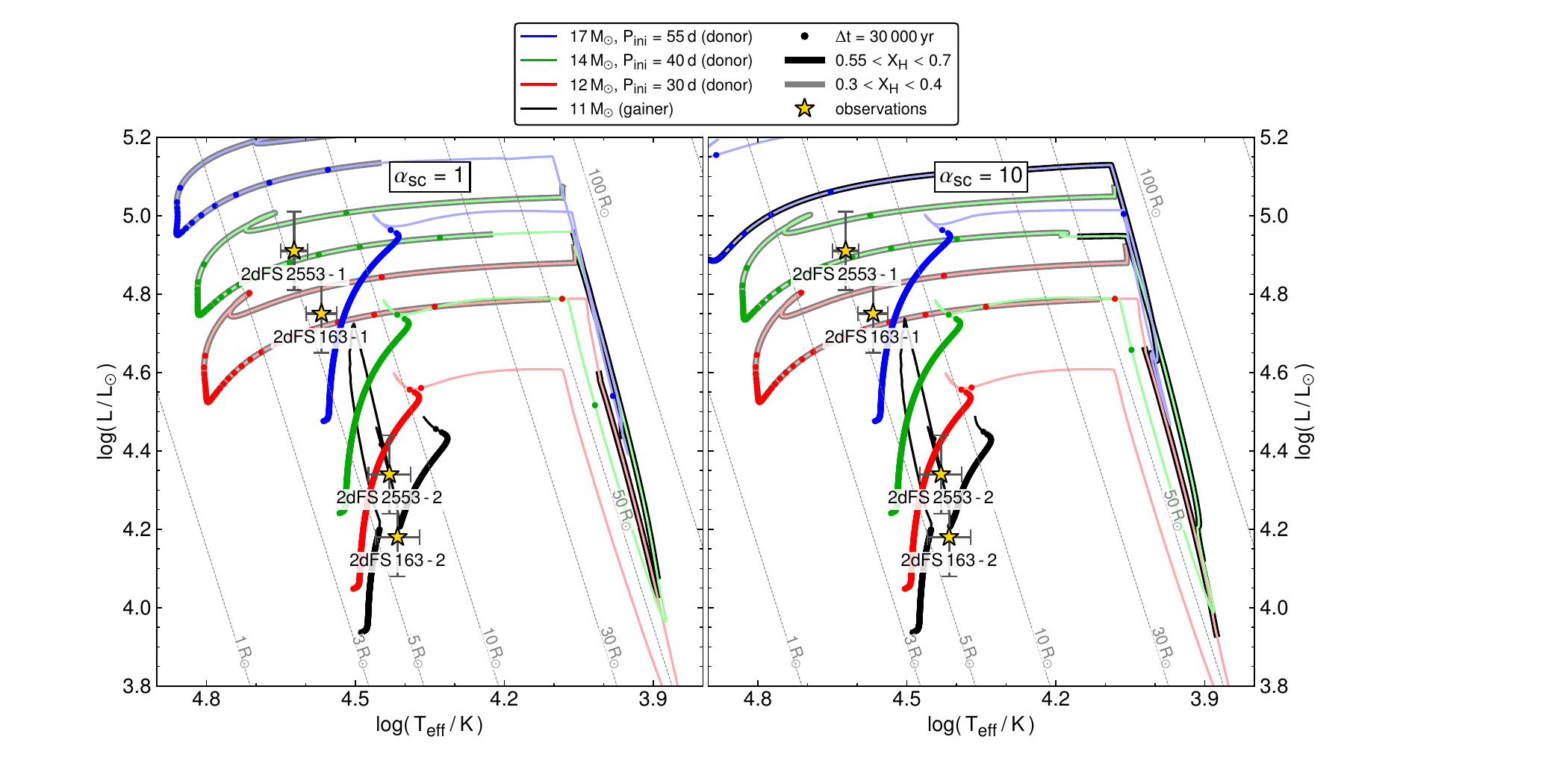}
        \caption{HRD showing possible evolutionary pathways of the primary and secondary components for two different sets of semiconvection mixing efficiencies. The donor tracks with initial masses $12\,M_\odot$, $14\,M_\odot$, and $17\,M_\odot$ are shown as red, green, and blue solid lines. The initial orbital periods are chosen, such that the binary has a period of about $P_\mathrm{model}=\SI{95\pm5}{d}$ after the mass-transfer event. The surface hydrogen abundances of our models that are consistent with the one measured for 2dFS~2553 is highlighted by a thick black contour and the one measured for 2dFS~163 as a thick gray contour. As the tracks of the secondary models do not differ drastically, only for clarity the track of the mass gainer of the system with the $17\,M_\odot$ donor is shown as a solid black line. To highlight phases where the stellar models spend most of their time, we over-plotted the tracks with dots, equally spaced in time steps of $\Delta t=\SI{30000}{yr}$. The observed positions of the primary and secondary components of the SMC binaries are shown as yellow stars.}
        \label{fig:MESA}
    \end{figure*}

In this paper, we discovered three systems with stripped-star primaries and Be/Oe companions. The stripped-star nature of primaries is revealed by their exceptionally low spectroscopic masses (Fig.~\ref{fig:ML}) and high nitrogen enrichment (Fig.~\ref{fig:CNO}). Their mass transfer origin is supported by fast rotation rates of companions, likely indicative of a recent mass and angular momentum accretion. Interestingly, the discovered stripped stars are all of O/B spectral types and with surface temperatures characteristic for MS stars, i.e., cooler and larger than the canonical He-burn sequence that stretches from subdwarfs to WR stars. Similar stars have previously been referred to as either bloated/puffed-up stripped stars \citep[e.g.,][]{Bodensteiner2020HR6819} or partially-stripped stars \citep[e.g.,][]{Ramachandran2023}, signaling that it is the remaining (unstripped) hydrogen envelope that is the cause of larger sizes and cooler surface temperatures compared to ``naked'' helium stars. If the remaining envelope is particularly massive, such a star may remain of O/B type for the entire core-He burning phase, i.e., $\sim$10\% of its total lifetime \citep{Klencki2022}. Otherwise, a puffed-up stripped star is a transitory stage lasting for $\sim$1\% the total lifetime \citep{DuttaKlencki2023} as the envelope is further depleted and the star contracts to become a hot and UV-bright stripped star \citep[][see also tracks in Fig.~\ref{fig:MESA}]{gotberg_2017_ionizing,Yungelson2024}. It may later puff-up again following the end of core-He burning, during a phase of re-expansion lasting for $\sim$0.1\% the total lifetime \citep{Laplace2020}. Finally, an OB-type stripped star could be a result of a stellar merger during which most of the envelope has been dynamically ejected, as proposed for the single star {\ensuremath{\gamma}} Columbae by \citet{Irrgang2022}.

Here, we model the newly discovered systems in the framework of binary evolution as products of stable mass transfer using the MESA stellar evolution code \citep{Paxton2011,Paxton2013,Paxton2015,Paxton2018,Paxton2019}. We focus on two SMC binaries, 2dFS~163 and 2dFS~2553, and exclude the LMC system, \mbox{Sk\,-71$^{\circ}$\,35} in which the mass transfer might currently still be ongoing. \mbox{Sk\,-71$^{\circ}$\,35} is also an ellipsoidal variable, which offers the possibility for a much more accurate determination of orbital and stellar parameters and tailored detailed modeling in a future work. Comparing 2dFS~163 and 2dFS~2553, we focus in particular on the surface H abundance.

While both primaries have roughly similar temperatures ($\sim\SI{40}{kK}$) and comparable luminosities ($\log(L/L_\odot)=4.75$ and $\log(L/L_\odot)=4.91$) they have both very distinct surface hydrogen abundances ($X_\mathrm{H}=0.33$ and $X_\mathrm{H}=0.6$). For the system with the hydrogen-rich primary, 2dFS~2553, the orbital period is known, and the Roche Lobe filling factor indicates that the system is detached.

As starting parameters for binary models, we set the initial masses as well as the initial orbital period of the system. For the initial primary masses, we calculate tracks with $M_\mathrm{ini,\,1}=12\,M_\odot, 14\,M_\odot$, and $17\,M_\odot$, corresponding to the observed luminosity range. For the secondary, we fixed the initial mass to $M_\mathrm{ini,\,2}=11\,M_\odot$, which is close to the masses of fast-rotating Be stars in 2dFS~163 and 2dFS~2553. Since the period of 2dFS~2553 is known to be about $P_\mathrm{orb,\,2dFS\,2553}=\SI{93.6}{d}$, we choose our initial periods such that the post mass transfer period is about $P_\mathrm{model}=\SI{95\pm5}{d}$. To reduce the free parameter space, we assume that the binary is initially tidally synchronized, which fixes the initial rotation of the stars. 
Semiconvective mixing is accounted for following \citet{langer1983}. It is one of the processes that will affect the chemical profile of deep envelope layers and, consequently, the surface abundances of stripped stars.
\citet{Schootemeijer2019} suggested a semiconvection efficiency of $\alpha_\mathrm{sc}\gtrsim 1$ to reproduce the observed blue/red supergiant ratio of the SMC. Motivated by this, we explore two sets of models: one with $\alpha_\mathrm{sc}=1$ and one with $\alpha_\mathrm{sc}=10$. Notably, even higher semiconvection values can lead to partially stripped stars with particularly massive residual envelopes and core-He burning lifetimes \citep{Klencki2022}.  
Further details on the assumed input physics within our models are provided in Appendix~\ref{app:MESA}.

Our two sets of binary models are compared with the observed primaries and secondaries are shown in Fig.\ref{fig:MESA}. In both the $\alpha_\mathrm{sc}=1$ and $\alpha_\mathrm{sc}=10$ cases,  the HRD positions of the partially stripped stars in 2dFS~163 and 2dFS~2553 overlap with the donor tracks of initial masses of $12\,M_\odot$ and $14\,M_\odot$, respectively, observed shortly after the end of mass transfer. The corresponding masses of the stripped donors in models are $4.5\,M_\odot$ and $5.6\,M_\odot$, with good agreement with the determined spectroscopic masses of primaries in 2dFS~163 and 2dFS~2553. The predicted surface H abundances are at around $X_\mathrm{H}\approx0.3$.  
While this is in agreement with the primary in 2dFS~163 ($X_\mathrm{H}=0.33$), it does not fit with the measurement for the primary of 2dFS~2553 ($X_\mathrm{H}=0.6$). 
This tension is partly reduced in models with $\alpha_\mathrm{sc}=10$, which tend to predict higher H abundance and would be consistent with $X_\mathrm{H}\approx0.6$ for a slightly more massive $17\,M_\odot$ donor.

The discrepancy in $X_\mathrm{H}$ may indicate that stellar evolution models are failing to accurately reproduce the abundance profile of deep envelope layers, which in turn affect the exact detachment point and the surface composition of stripped stars formed via mass transfer. Besides semiconvection, multiple other agents are at play, such as rotational mixing, convective-boundary mixing, and gravity-wave induced mixing \citep{Georgy2013,Kaiser2020,Klencki2020,Johnston2021,Pedersen2022}. The matter is further complicated if the surface abundances of stars in a binary that has only very recently detached from mass transfer are influenced by back-accretion of even a small amount of leftover matter from the circumbinary medium.
 Interestingly, a surprisingly high surface hydrogen of $X_\mathrm{H} \gtrsim 0.55-0.6$ has been measured in most of the known puffed-up/partially stripped stars originating from massive primaries \citep{Ramachandran2023,Villasenor2023,Pauli2022b}, suggesting this to be the norm rather than an exception. From this perspective, 2dFS~163 with $X_\mathrm{H}=0.33$ stands out as the first of these objects with a significantly reduced surface hydrogen. This might suggest a slightly different evolutionary pathway leading to more efficient envelope stripping, for instance case A + AB mass transfer rather than case B evolution (to be resolved with follow-up data constraining the orbit). It might also be that 2dFS~163 is observed during a later evolutionary stage of post-He burning re-expansion \citep{Laplace2020}. Although shorter-lived, puffed-up stripped stars of this stage are expected to be more He enhanced and be somewhat closer to the Eddington limit, as is the case for 2dFS~163 (Fig.~\ref{fig:ML}).

    It is worth noting, that due to the usage of the low mass-loss rates of \citet{Vink2017} during the post-mass transfer phase, our models never loose their full hydrogen-rich envelope and they re-expand again following core-He depletion. However, this might only be the case for models undergoing mass transfer after the main sequence evolution (Case B), as the removal of the hydrogen-rich envelope in the mass-transfer events during the main sequence (Case A) is more efficient.

    The newly discovered systems contribute to the population of post-interaction binaries with intermediate-mass stripped stars. A handful of such systems is now known in the mass range of $\gtrsim12\,M_\odot$ (pre-interaction), i.e., firmly in the mass regime of neutron star progenitors. In their future evolution, the stripped primaries will explode as type Ib/c or IIb supernovae. If the orbit survives the explosion, the systems will likely evolve to become Be X-ray binaries. Those with wide orbits (hundreds of days) may further go through and survive a common-envelope phase \citep{Tauris2017}. Massive post-interaction binaries reported in this work are thus a much needed anchor point for understanding mass transfer evolution in the formation channels of X-ray binaries and gravitational-wave sources. This growing class of objects includes cases in which conservative mass accretion is required, e.g., nearly 100\% in \citealt{Villasenor2023} and at least 30\% in \citealt{Ramachandran2023}. While the newly discovered \mbox{Sk\,-71$^{\circ}$\,35} is not part of our binary model analysis, the inferred spectroscopic masses $M_1 = 7.8^{+3.6}_{-3}\,M_\odot$ and $M_2 = 35.7^{+18}_{-14}\,M_\odot$ suggest that a significant fraction of the transferred mass has been accreted. On the other hand, both 2dFS~163 and 2dFS~2553 reported here can be reproduced with nearly fully non-conservative mass transfer and agree well with the accretion efficiency being limited by critical rotation (as adopted in the models in Fig.~\ref{fig:MESA}).  Notably, the secondaries in models with $\alpha_\mathrm{sc}=1$ rotate with rotation velocities close to their breakup velocity, while secondaries in models with $\alpha_\mathrm{sc}=10$ slow down to only about $~80\%$ of their breakup velocity after detachment. The latter is in better agreement with observations.

\begin{figure}
    \centering
    \includegraphics[width=1.0\linewidth]{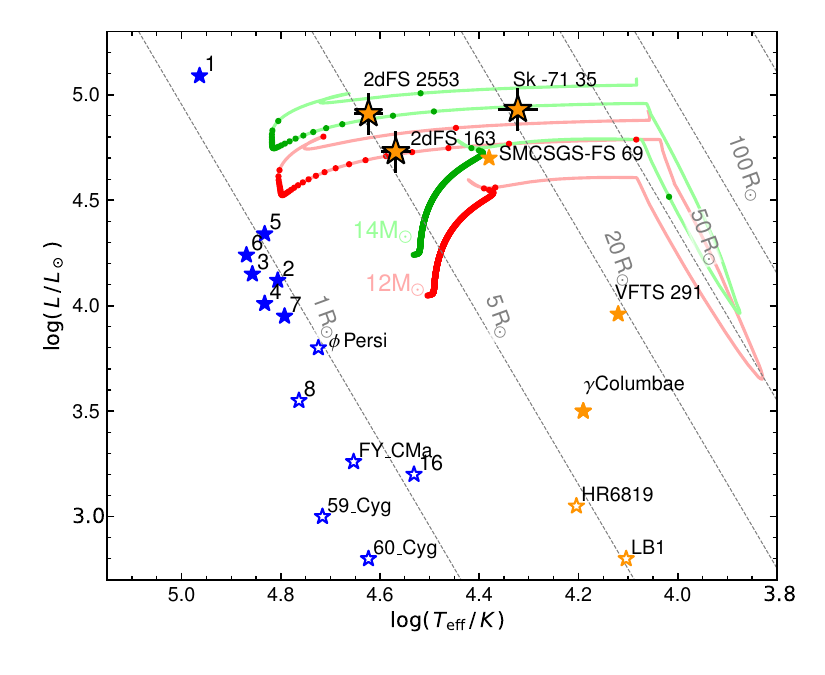}
    \caption{HRD for stripped stars including objects discovered in this work and those from literature \citep{Gotberg2023,Ramachandran2023,Villasenor2023,Irrgang2022,Bodensteiner2020HR6819,Shenar2020lb1,Schootemeijer2018,Wang2021}. Orange star symbols represent partially stripped stars, while blue are compact stripped He stars.  The open star symbols denote low-mass stars in the subdwarf regime, while the filled ones correspond to intermediate masses. Blue stars with numbers 1-8, and 16 are stripped stars located in the Magellanic Clouds \citep{Gotberg2023}, VFTS\,291 is part of the LMC, and SMCSGS-FS\,69 is in the SMC. The rest of the samples are Galactic sources. The evolutionary tracks in the background have the same meaning as in Fig.\,\ref{fig:MESA}.}
    \label{fig:hrdstrip}
\end{figure}

The stripped star binaries detected to date may still be just the tip of the iceberg (see HRD in Fig.\,\ref{fig:hrdstrip}). 
\citet{DuttaKlencki2023} estimate that at least $\sim$$0.5-1\%$ of O/B/A stars may in fact be post-interaction systems with puffed-up stripped stars originating from primaries with $M_\text{ini} >3\,M_{\odot}$. This corresponds to $\sim100$ systems in the SMC alone and about 5-10 in the luminosity range $4.6 < \log (L/L_{\odot}) < 5.0$. Combining the 2 new SMC systems reported here as well as SMCSGS-FS\,69 from \citet{Ramachandran2023}, we already have 3 post-interaction binaries in that luminosity range. While this number is currently consistent with the prediction, there is a good chance that the underlying population is at least a few times larger as the search so far has been limited to only a small fraction of the overall population of $\sim$$1000$ OB stars in the SMC. If that turns out to be the case, binary evolution models may need to be refined to at least sometimes produce long-lived partially stripped stars: whether through inefficient envelope stripping \citep{Klencki2022} or an unknown mechanism causing premature detachment from mass transfer, stellar mergers with substantial mass loss \citep{Irrgang2022}, or something else entirely.

On the other hand, binary evolutionary models predict a large fraction of stripped stars to be on the He ZAMS, spanning almost 10\% of their total lifetime \citep{Gotberg2023,Yungelson2024}. Thus, we would expect to observe at least 10 times more hot and compact stripped He stars than puffed-up stripped stars. As shown in Fig.\,\ref{fig:hrdstrip}, we lack observations of such stripped stars on the He ZAMS compared to puffed-up stripped stars in the higher luminosity range of $\log (L/L_{\odot}) > 4.5$.
Our newly discovered luminous, partially stripped stars were initially chosen as potential subjects for further examination due to the existence of a Be/Oe companion. The Be phenomenon is expected to be transient, potentially occurring more frequently in stripped-star binaries that have recently undergone mass transfer while being less common in older, hot stripped star systems. As a result, we anticipate detecting more binary systems with stripped stars that do not necessarily have a Be/Oe companion in the XshootU sample and in the Magellanic Clouds in general, with future studies involving multi-epoch data and with detailed orbital analysis. 
    In the lower luminosity range $3.5 < \log (L/L_{\odot}) < 4.5$, we do have more observations of hot and compact stripped stars relative to puffed-up stripped stars (e.g., in the Magellanic Clouds the current observed ratio in this regime is 7:1 and thus more in line with evolution model expectations). 

     Whether or not a considerable number of stripped stars spend a large amount of their post-MS lifetime as a puffed-up O/B star has a significant impact on the role of stripped stars as sources of ionizing radiation. Both the primaries and the secondaries are relevant sources of \ion{H}{i} ionizing photons, with their fluxes being similar to O/B MS stars of the same luminosity. However, the cooler temperatures of puffed-up stripped primaries essentially prevent them from being considerable sources of \ion{He}{II} ionizing radiation (cf.\ Tables \ref{table:parameters}, \ref{table:parameters2}, \ref{table:parameters3}). Even the earliest star in our sample, 2dFS~2553, contributes only $\sim$$10^{43}$ photons per second, much less than what would be expected for stripped stars on the He main sequence \citep[e.g.,][]{Gotberg2018}. Presently, population synthesis models that aim to include binary-stripped stars below the WR regime rely on atmosphere models representing He stars residing in a hot, compact stage \citep[e.g.,][]{Gotberg2019,Lecroq2024}. Depending on the total number of detections and the derived parameters of stripped stars, this treatment might need to undergo considerable revision.

\section{Summary and Conclusions}
\label{summary}

Using UV and optical spectroscopy, we discovered three partially stripped star + Oe/Be binaries in the XShootU sample. Two of these systems are located in the SMC and one is found in the LMC. We identified distinct spectral characteristics of both the partially stripped primary star, which exhibits comparably narrow lines and prominent nitrogen lines and the secondary star, which displays broad lines and disk emission features. By combining the XShootU data with additional archival spectra, we have confirmed RV variations and observed components moving in anti-phase in all three systems. 
We carried out a composite spectral analysis of the multi-wavelength spectra using the PoWR model atmospheres. We determined the period of the two systems and performed orbital analysis. The comprehensive analyses lead us to the following conclusions:

\begin{itemize}
\item The detected binaries contain partially stripped stars of masses in the range of $\approx 4-8\,M_{\odot}$. The spectroscopic masses are found to be consistent with evolutionary masses and orbital masses, within uncertainty.   Our stripped stars are thus of intermediate mass, between the ranges of WR and subdwarf masses, where there is so far a noticeable absence of observations. 
\item In all three systems, the Be/Oe secondaries are found to be significantly more massive than their partially stripped companions, showing mass ratios of $q \gtrsim 2$.
\item The partially stripped stars in 2dFS\,163 and 2dFS\,2553 are as hot as early O-type stars and have radii similar to those on the ZAMS. On the other hand, \mbox{Sk\,-71$^{\circ}$\,35} is a bloated stripped star with spectral features resembling a B supergiant. Both the light curve and spectroscopic analysis indicate that this system may be currently undergoing mass transfer.
\item The partially stripped stars in our sample are over-luminous for their corresponding stellar masses, which is consistent with the luminosities during core He-burning.
\item The surfaces of all three partially stripped stars clearly show signs of substantial nitrogen enrichment, along with reduced carbon and oxygen abundances. In addition, 2dFS\,163 exhibits significant indications of helium enrichment.

\item We find no evidence of an increased ionizing flux contribution (H and \ion{He}{ii}) in our sample of partially stripped stars compared to OB stars of similar luminosity or temperatures. 

\item The wind mass-loss rates of the newly found partially stripped stars are determined to be on the order of $10^{-7}\,M_\odot\,\mathrm{yr}^{-1}$, which is more than ten times higher than that of (main sequence) OB stars with the same luminosity. The current mass-loss recipes commonly employed in evolutionary models to characterize this phase have been found to be significantly under- or overestimated by an order of magnitude. With considerable scatter, the \citet{Vink2017} description provides the right order of magnitude in $\dot{M}$ for their optically thin winds, at least for our observed sample.

\item The $\varv_\infty$ to $\varv_{\mathrm{esc},\Gamma}$ ratio for stripped stars is in agreement with those derived for other OB stars, but their $\varv_\infty$ is much lower than observed for single OB stars with similar effective temperatures.

\item Binary evolution models can explain the current observed temperatures, luminosities, and stellar masses of the primaries in 2dFS\,163 and 2dFS\,2553. These models suggest that both primaries are core-helium burning. Their distinct surface hydrogen abundances might indicate that more efficient mixing within the stellar models is needed, which would change our standard picture of intermediate-mass binary evolution. Further follow-up in-depth studies are required to investigate the potential consequences.

\item The detection of multiple objects with parameters traditionally associated to a short-living transition stage provides a potential challenge to current binary evolution models. While the current number of systems is still in line with predictions, further detections in larger binary-focused surveys such as BLOeM could significantly increase the number of systems. A considerable number of partially stripped stars hiding among the presumed OB main sequence parameter space would also have consequences for the integrated parameters from population synthesis models such as their \ion{He}{ii} ionizing fluxes.

\item Our newly discovered partially stripped stars are firmly in the mass regime to produce neutron stars and stripped-envelope supernovae. If these systems survive the explosion, they will likely evolve into Be X-ray binaries.

\end{itemize} 

\begin{acknowledgements}
The spectral plots in this work were created with WRplot, developed by W.-R. Hamann. We thank the anonymous referee for providing constructive comments that helped improve the paper.
VR, AACS, and MBP are supported by the Deutsche Forschungsgemeinschaft (DFG - German Research Foundation) in the form of an Emmy Noether Research Group -- Project-ID 445674056 (SA4064/1-1, PI Sander) and funding from the Federal Ministry of Education and Research (BMBF) and the Baden-Württemberg Ministry of Science as part of the Excellence Strategy of the German Federal and State Governments. 
JSV is supported by STFC funding under grant number ST/V000233/1. RK acknowledges financial support via the Heisenberg Research Grant funded by the Deutsche Forschungsgemeinschaft (DFG, German Research Foundation) under grant no.~KU 2849/9, project no.~445783058.
We thank C.J.\ Evans for providing the reduced 2dF spectra. We acknowledge the use of TESS High Level Science Products (HLSP) produced by the Quick-Look Pipeline (QLP) at the TESS Science Office at MIT, which are publicly available from the Mikulski Archive for Space Telescopes (MAST). Funding for the TESS mission is provided by NASA's Science Mission directorate.
   
\end{acknowledgements}

\bibliographystyle{aa} 
\bibliography{main} 

\appendix

\section{Radial velocity measurements for 2dFS\,163}
\label{appendix:2dfs163}
\update{For \mbox{2dFS\,163}, we only have 3 epochs of available spectra to measure the radial velocities. For the primary (i.e., the slowly rotating stripped star) we derived the RVs by fitting Gaussians to the associated narrow metal lines in the optical and UV spectra.
In the Xshooter spectrum, we used the \ion{N}{iii\,$\lambda\lambda$4634--4642} emission lines and absorption lines such as \ion{N}{iv\,$\lambda\lambda$3460-3485} and \ion{He}{i\,$\lambda$4713}
for this purpose. Afterwards, we cross-correlated both X-shooter and 2dF spectrum to get the corresponding RV for 2dF. For this, we focused on nitrogen lines, and we got a $\delta_{\mathrm{RV}}=28$\,km\,s$^{-1}$. In the UV spectrum, we used narrow lines such as \ion{N}{iii\,$\lambda$1183} and \ion{C}{iv\,$\lambda$1169} to determine the RV of the stripped star. 
To estimate the secondary's RVs, we fitted the synthetic model spectra of the primary and secondary to the observations while fixing the primary's RV to the values derived with the method described above. 
}

\begin{table}
    \centering
    \caption{\update{RV estimates for \mbox{2dFS\,163}. Note that all spectra are corrected for the barycentric motion but not for the systemic velocity of the SMC.}}
    \renewcommand{\arraystretch}{1.6}
    \begin{tabular}{cccccc}
        		\hline 
		\hline
		\vspace{0.1cm}
     Instrument & MJD &  RV1 &$\delta$RV1& RV2 &$\delta$RV2 \\
     \hline
   2dF & 51083   & 170 &20 & - & -\\
 X-shooter  & 59150.0735 & 142 &10 &  105 & 30\\
HST/COS & 59131.2654  & 185 &10 & 120 & 30\\

    \hline
    \end{tabular}
    \label{tab:RV163}
\end{table}

\section{Orbital analysis of 2dFS\,2553}
\label{appendix:2dfs2553}

\subsection{Radial velocity measurements}
\label{appendix:RV}
For this binary, we are in the lucky position to have sufficient (>\,3) spectra that can be used to perform an orbital analysis, yielding an estimate for possible periods and the mass ratio. For the seven spectra we have at hand, we estimated the RVs following different approaches: For the primary (i.e. the slowly rotating partially stripped star) we derived the RVs by fitting Gaussians to the associated narrow metal lines in the optical and UV spectra. For the secondary (the fast-rotating Be star) we also fitted Gaussians to the broad \hei lines in the optical spectra. In the UV spectra the primary is dominating the spectral appearance and the secondary's contribution can only be seen in the \ion{C}{iii$\,\lambda1175$} line. To estimate the secondary's RVs, we fitted the synthetic model spectra of the primary and secondary to this line while keeping the primary's RV fixed to the values derived from the other lines. In addition to our estimated RV measurements,  \cite{Lamb2016} also analyzed this star and reported a radial velocity shift for the lines associated with our primary star. Since their spectrum is not available in public archives, we adapt their reported RV for our primary. The derived RVs are listed in Table\,\ref{tab:RV}.

\subsection{Constraining ephemerides with \textit{The Joker}}
\label{appendix:Joker}
At first, we employed the \textit{The Joker} \citep{Price-Whelan2017}, a custom Monte Carlo sampler designed to derive possible ephemerides for sparse RV datasets. Unfortunately, it is only applicable to the RV of one star and, hence, does not yield and information about parameters like the mass ratio. In order to sample quickly a large parameter space, \textit{The Joker} performs at first a rejection sampling analysis by using prior probability density functions (PDFs) and yields multimodal PDFs of the ephemerides. Within this work, we use the default PDFs as recommended by the manual. As input a log-uniform priors in the range of $\SIrange{0.5}{700}{d}$. From this first quick sampling, \textit{The Joker} reports possible periods to be around the order of a few days ($\sim\SI{7}{d}$, up to $\SI{300}{d}$, with the most likely solution to be around $\SI{95}{d}$. Note, that the most extreme solutions (i.e.\ the shortest and longest periods) favor eccentric orbits ($e>0.2$), while the most likely solution around $\SI{95}{d}$ favors circular orbits. 

After the rejection sampling, a standard MCMC method is used to sample around the most likely solution to
fully explore the posterior PDF. Following this procedure, we find that the RVs can be modeled the best with a period of $P=93.67^{+0.05}_{-0.07}\,\si{d}$ and a circular orbit ($e=0^{+0.09}$).

\subsection{Full orbital analysis with PHOEBE}
\label{appendix:PHOEBE}
As a second step, we employ the Physics of Eclipsing Binaries (PHEOBE) code \citep[version 2.3,][]{Prsa2005,Prsa2016,Horvat2018,Jones2020,Conroy2020}  to determine all orbital parameters consistently from the primary's and secondary's RVs. For our fitting efforts, we use the built-in option of the MCMC sampler emcee (Foreman-Mackey et al. 2013), allowing us to probe a large parameter space and obtain realistic error margins on the individual orbital parameters.

\begin{table}
    \centering
    \caption{RV estimates for 2dFS\,2553. Note that all spectra are corrected for the barycentric motion but not for the systemic velocity of the SMC.}
    \renewcommand{\arraystretch}{1.6}
    \begin{tabular}{cccccc}
        		\hline 
		\hline
		\vspace{0.1cm}
     Instrument & MJD &  RV1 &$\delta$RV1& RV2 &$\delta$RV2 \\
     \hline
    2dF & 51083 &  260  &  20  &  150 & 40   \\
    IMACS$^{(a)}$ & 55069 &   130&  20  &  --- & ---  \\
    HST/STIS & 59024.264 &   210 &  10  & 200 & 40   \\
    HST/COS & 59744.9111 &  80 &  10 &   190&  30  \\
    HST/COS & 59761.6678 &  155&   10 &  190&  30  \\
    X-shooter  & 59161.1254& 127&   10 &  215&  20  \\
    X-shooter & 59805.2514&  214&   10 &  143&  20  \\
    HST/COS & 60083.8085&  210&   10 &  120&  30  \\   
    \hline
    \end{tabular}
    \begin{minipage}{0.95\linewidth}
                \ignorespaces 
                \textbf{Notes.} $^{(a)}$ Value adopted from \cite{Lamb2016}\\
            \end{minipage}
    \label{tab:RV}
\end{table}
\begin{figure}[tb]
\footnotesize
    \centering
    \includegraphics[width=\hsize]{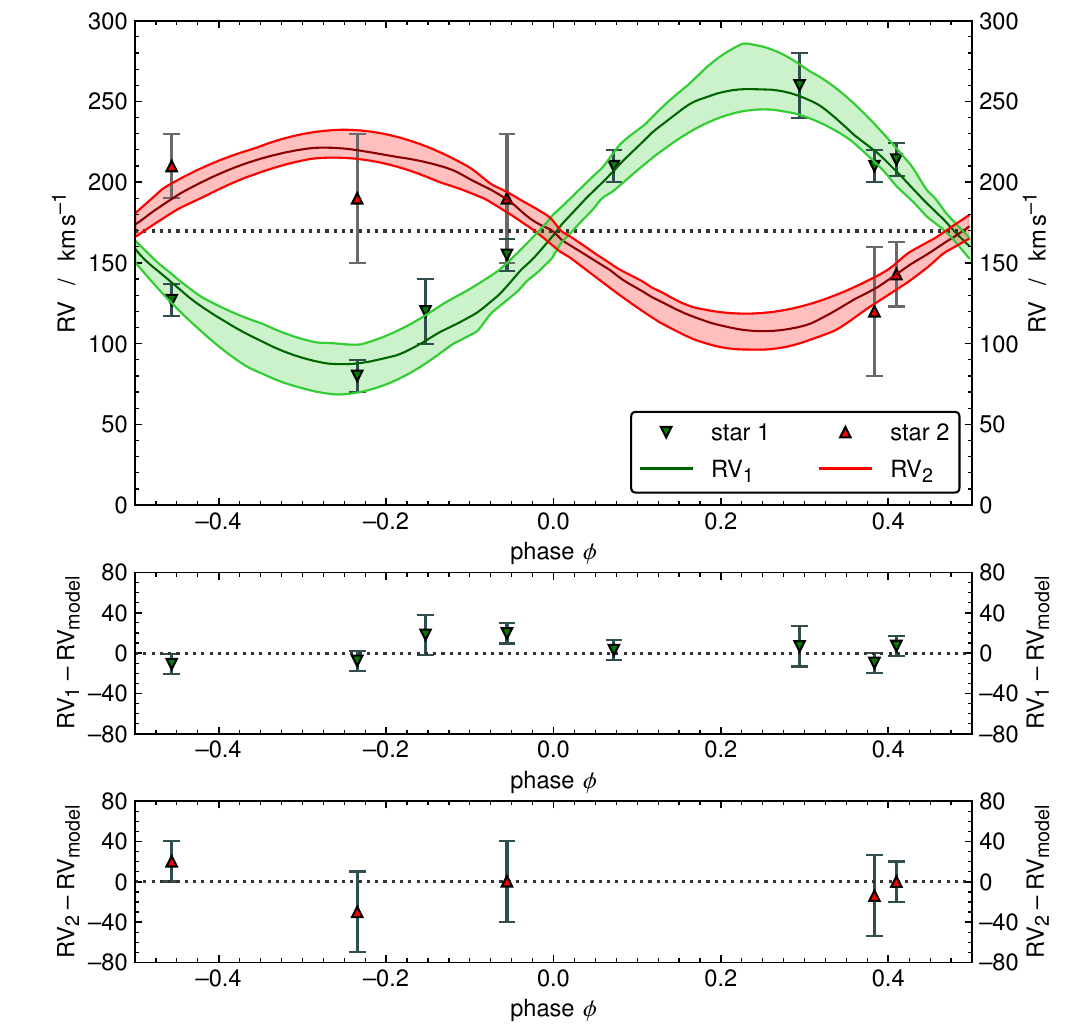}
    \caption{\textit{Upper panel:} Observed (triangles) and synthetic (solid lines) RV curves of the primary (green) and secondary (red) components in 2dFS\,2553, as obtained with the PHOEBE code. The dashed black line indicates the systemic velocity $\varv_\gamma$. \textit{Middle panel:} Residuals of the primary's RVs after subtraction of the model. \textit{Lower panel:} Residuals of the secondary's RVs after subtraction of the model.}
    \label{fig:phoebe_RVs}
\end{figure}

\begin{table}[tb]
        \footnotesize
        \centering
        \caption{Orbital parameters for 2dFS\,2553 derived with the PHOEBE code.}
        \begin{tabular}{ccc}
                \hline\hline \rule{0cm}{2.8ex}
                \rule{0cm}{2.8ex}parameter & value & unit\\ 
                \hline
                \rule{0cm}{2.8ex}$P_\mathrm{orb}$ & $93.66^{+0.07}_{-0.08}$ & $[\mathrm{d}]$\\ 
                \rule{0cm}{2.6ex}$\omega_0$ & $134^{+107}_{-98}$ & $[\mathrm{^\circ}]$\\ 
                \rule{0cm}{2.6ex}$t_0$ & $54929.38^{+7.5}_{-4.5}$ & $[\mathrm{MJD}]$\\ 
                \rule{0cm}{2.6ex}$\varv_\gamma$ & $169.78^{+2.75}_{-6.15}$ & $[\mathrm{km\,s^{-1}}]$\\ 
                \rule{0cm}{2.4ex}$e$ & {$0.03^{+0.14}_{-0.03}$} &\\ 
                \rule{0cm}{2.6ex}$q_\mathrm{orb}$ & $1.30^{+0.55}_{-0.24}$ &\\ 
                \rule{0cm}{2.6ex}$a\,\sin\,i$ & $211.1^{+51.8}_{-20.2}$ & $[R_\odot]$\vspace{1ex}\\ 
                \hline
                \rule{0cm}{2.8ex}$i_\mathrm{orb}\,^{(a)}$ & $70$ & $[\mathrm{^\circ}]$\\ 
                \rule{0cm}{2.6ex}$a\,^{(b)}$ & $224.7^{+21.5}_{-55.1}$ & $[{R_\odot}]$\vspace{1ex}\\ 
                \hline 
        \end{tabular}
                \rule{0cm}{2.8ex}%
                \tablefoot{
                \ignorespaces
                $^{(a)}$ Calibrated such that the orbital and spectroscopic masses agree within their uncertainties. $^{(b)}$ Calculated using $i_\mathrm{orb}$. %
                }
        \label{table:phoebe_results1}
    \end{table}
    
    \begin{table}[tb]
    \footnotesize
        \centering
        \caption{Orbital parameters of the individual components in 2dFS\,2553 obtained using the PHOEBE code.}
        \begin{tabular}{cccc}
                \hline\hline \rule{0cm}{2.4ex}
                \rule{0cm}{2.8ex}parameter & primary & secondary & unit\\ 
                \hline
                \rule{0cm}{2.8ex}$K$ & $65^{+3}_{-3}$ & $50^{+4}_{-4}$ & $[\mathrm{km\,s^{-1}}]$\\ 
                \rule{0cm}{2.6ex}$M_\mathrm{orb}\,\sin\,^3\,i$ & $6.26^{+4.84}_{-1.91}$ & $8.14^{+6.17}_{-2.43}$& $[{M_\odot}]$\vspace{1ex}\\ 
                \hline
                \rule{0cm}{2.8ex}$M_\mathrm{orb}\,^{(a)}$ & $7.5^{+5.8}_{-2.3}$ & $9.8^{+7.4}_{-2.9}$ & $[{M_\odot}]$\\ 
                \rule{0cm}{2.6ex}$R_\mathrm{RL}^{\,\,\,\,\,\,\,\,\,(a)}$ & $80^{+18}_{-8}$ & $90^{+24}_{-8}$&$[{R_\odot}]$\vspace{1ex}\\ 
                \hline 
        \end{tabular}
        \rule{0cm}{2.8ex}%
                \tablefoot{
                \ignorespaces
                $^{(a)}$ Calculated using $i_\mathrm{orb}$ (see Table~\ref{table:phoebe_results1}).  %
                }
        \label{table:phoebe_results2}
    \end{table}

To derive the best fitting model, we used as free parameters the orbital period of the star $P_\mathrm{orb}$, the eccentricity $e$, the projected orbital separation $a\sin\,i$, the argument 
of the periastron $\omega$, the time at which the primary component in our orbit is at superior 
conjunction $t_0$, the mass ratio $q$, and the systemic velocity $\varv_\gamma$. The prior distributions for  $P_\mathrm{orb}$, $a\sin\,i$, $t_0$, and $\varv_\gamma$ are assumed to have
Gaussian shape, centered at the values found by \textit{The Joker}. For the remaining parameters, we assume starting parameters of $e=0$, $\omega=0$, and $q=2$ (the spectroscopic mass ratio) 
and assume that they are uniformly distributed over the ranges $e=\SIrange{0}{0.4}{}$, $\omega=\SIrange{0}{2}{}\pi\,\si{rad}$, and $q=\SIrange{0.33}{3.0}{}$.

The best values we quote here are the ``mode'' values (i.e. the most populated solutions and, hence, the most probable one), and their corresponding error margins are given as one-sigma confidence intervals. Our best RV fit is shown in Fig.\,\ref{fig:phoebe_RVs} the best fitting values are listed in Tables\,\ref{table:phoebe_results1} and \ref{table:phoebe_results2}.

The PHOEBE results on the period and eccentricity are comparable with those derived by \textit{The Joker}, but favor a slightly eccentric solution but still consistent with a circular orbit in the error margins. The derived mass ratio is  $q_\mathrm{orb}=1.3$ a bit lower than the spectroscopically derived one $q=2^{+1.2}_{-0.96}$. However, both values agree within their respective uncertainties.

We tried to approximate the inclination by matching the projected masses $M_\mathrm{orb}\,\sin\,^3\,i$ to the spectroscopic masses, which yields $i\approx70^\circ$. The spectroscopically derived projected rotation rate of the Be secondary $\varv_\mathrm{rot}\,\sin i = 300^{+100}_{-50}\,\si{km\,s^{-1}}$ would change to $\varv_\mathrm{rot}=319^{+106}_{-52}\,\si{km\,s^{-1}}$, which corresponds to  $\varv / \varv_{\mathrm {crit}} \approx0.63$.

\section{Orbital analysis of \mbox{Sk\,-71$^{\circ}$\,35}}
\label{appendix:sk7135}

\subsection{Radial velocity measurements}
\label{appendix:RV2}
For \mbox{Sk\,-71$^{\circ}$\,35}, we only have 3 epochs of available spectra to measure the radial velocities. For the primary (i.e. the slowly rotating partially stripped star) we derived the RVs by fitting Gaussians to the associated narrow metal lines in the optical and UV spectra. For the secondary (the fast-rotating Oe star) we also fitted Gaussians to the broad \hei and \heii lines in the optical spectra. Here especially the \ion{He}{ii\,$\lambda$4686} and \ion{He}{ii\,$\lambda$5412} are only coming from the secondary. 
In the UV spectra,  \ion{Si}{iv\,$\lambda\lambda$1393--1403} P\,Cygni profiles are visible from both primary and secondary and are well-separated. So we use this to estimate the RVs by fitting the synthetic model spectra of the primary and secondary to this line. In addition, we also use \ion{N}{v} lines in the UV to estimate the RV of the secondary.  The derived RVs for \mbox{Sk\,-71$^{\circ}$\,35} are listed in Table\,\ref{tab:RV2}.

\begin{table}
    \centering
    \caption{RV estimates for \mbox{Sk\,-71$^{\circ}$\,35}. Note that all spectra are corrected for the barycentric motion but not for the systemic velocity of the LMC.}
    \renewcommand{\arraystretch}{1.6}
    \begin{tabular}{cccccc}
        		\hline 
		\hline
		\vspace{0.1cm}
     Instrument & MJD &  RV1 &$\delta$RV1& RV2 &$\delta$RV2 \\
     \hline
    GIRAFFE & 57375.175   & 130 &10 & 280 & 30\\
 X-shooter  &59285.02995 & 345 &10 & 240 & 30\\
HST/COS &59689.9703  & 440 &20 & 230 & 30\\

    \hline
    \end{tabular}
    \label{tab:RV2}
\end{table}

\subsection{TESS light curve analysis}
\label{appendix:tess}

\begin{figure}[!hbt]
   \centering
   \includegraphics[width=8.5cm]{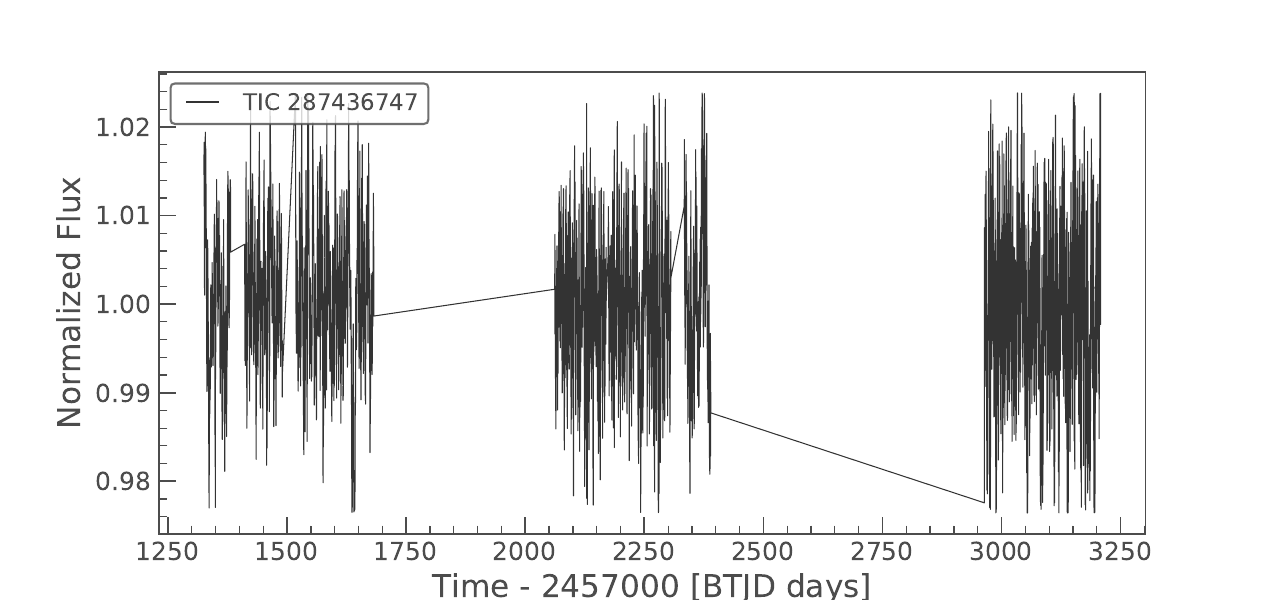}
    \includegraphics[width=8.5cm]{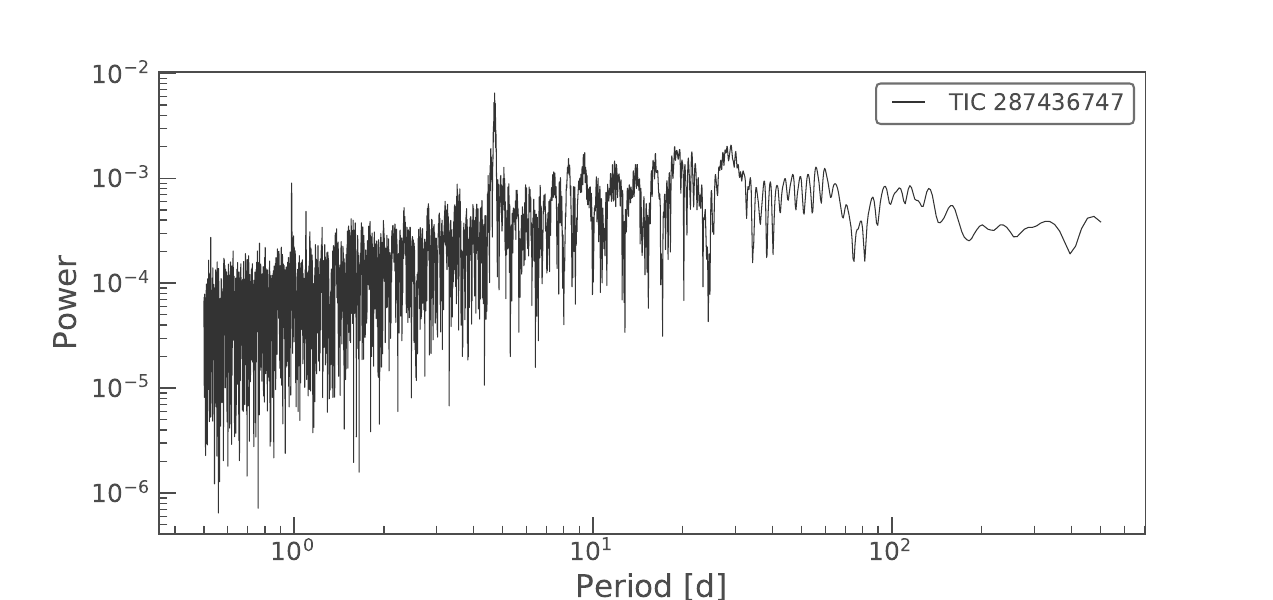}
     \includegraphics[width=8.5cm]{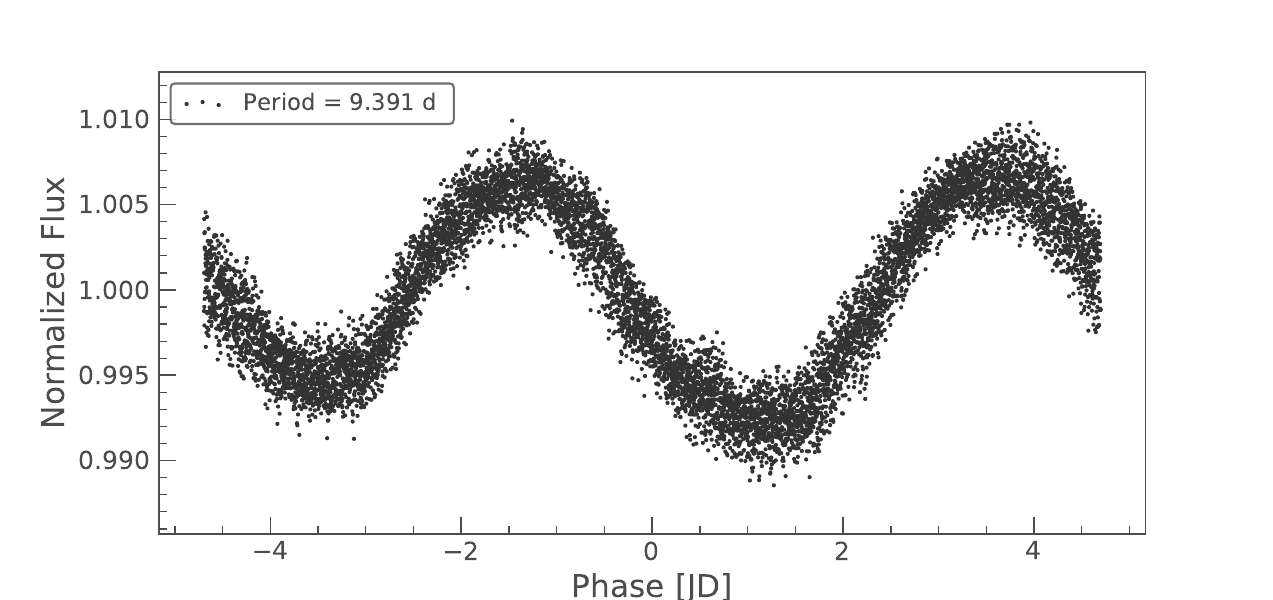}
      \caption{TESS light curves for \mbox{Sk\,-71$^{\circ}$\,35} taken during 2018 to 2023 (upper panel). The middle panels show the periodograms of the light curve, and the bottom panel includes the phase-folded light curve with a period of $\approx9.4$  days.}
         \label{LC}
\end{figure}

To determine the orbital period we make use of TESS light curves. We used the available 31 TESS light curves from the MIT Quick-Look Pipeline \citep[QLP,][]{Huang2020} taken from 2018 to 2023. Using the Lightkurve package \citep{Lightkurve2018}, we concatenate all light curves together, followed by removing outliers ($3\sigma$ limit) and normalizing all of them at the same time. The resulting light curve is shown in Fig.\,\ref{LC} (upper panel). We performed a period search and found a clear peak at $\approx4.7$ days corresponding to a binary orbital period of $\approx9.4$ days.

\subsection{Orbital analysis with \texttt{rvfit}}
\label{appendix:rvfit}
The RV measurements were analyzed using the \texttt{rvfit} code that models the RV curves for double-lined, single-lined binaries,
or exoplanets \citep{Iglesias-Marzoa2015}. To do this, it uses 
an Adaptive Simulated Annealing (ASA) algorithm \citep{Ingber1996}  which fits the following
seven Keplerian parameters:
 the orbital period of the star $P$, the eccentricity $e$, the argument 
of the periastron $\omega$, the time  of periastron passage $T_p$,  the systemic velocity $\gamma$,  and the amplitude of the radial velocity K1 for the primary and K2 for the secondary star.

\begin{figure}
    \centering
    \includegraphics[width=0.9\linewidth]{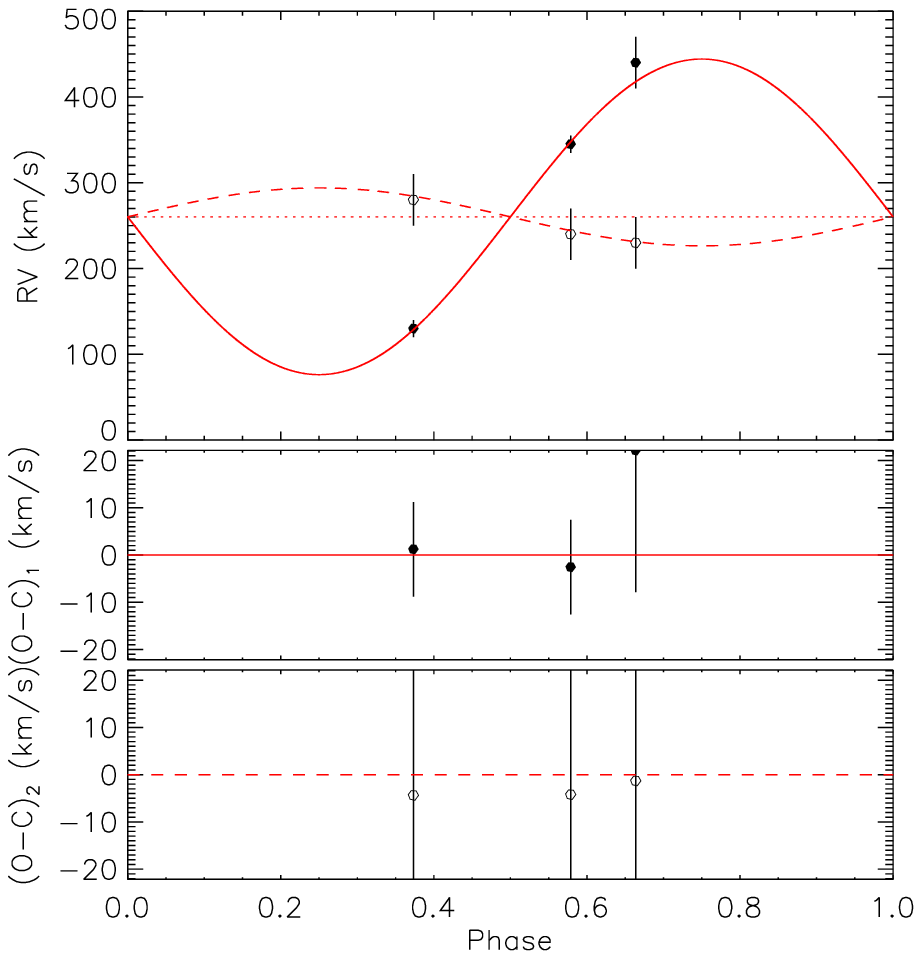}
    \caption{RV curve fit for primary (filled circles, solid curve) and secondary  (open circles, dashed curve)  components in \mbox{Sk\,-71$^{\circ}$\,35} obtained with  \texttt{rvfit}. The bottom two panels show the residuals of the fit for the primary and the secondary.}
    \label{fig:rvfit}
\end{figure}

The period of the system is measured from the TESS light curve. In the
analysis, we assume that the system has a circular
orbit. Hence, those parameters were fixed during further analysis.
The resulting  \texttt{rvfit} orbital solutions are given in Table\,\ref{table:rvfit_results}. The theoretical RV fits to the measurements are shown in Fig.\,\ref{fig:rvfit}. 

The mass ratio derived based on the orbital analysis is  $q_\mathrm{orb}=5.4$, higher than the spectroscopically derived one $q=4.6^{+3.1}_{-2.5}$. However, both values agree within their respective uncertainties.
We tried to approximate the inclination by matching the projected masses $M_\mathrm{orb}\,\sin\,^3\,i$ to the spectroscopic masses, which yields $i\approx37^\circ$. The corresponding orbital masses are $M_1\approx7.1\,M_{\odot},\,M_2\approx38.8\,M_{\odot} $ and the orbital separation is $67\,R_{\odot}$. The spectroscopically derived projected rotation rate $\varv_\mathrm{rot}\,\sin i \approx 250 \,\si{km\,s^{-1}}$ would change to $\varv_\mathrm{rot} \approx 415\,\si{km\,s^{-1}}$, which corresponds to  $\varv / \varv_{\mathrm {crit}} \approx0.67$.

\begin{table}[tb]
        \footnotesize
        \centering
        \caption{Orbital parameters derived  for \mbox{Sk\,-71$^{\circ}$\,35} with  with \texttt{rvfit}}
        \begin{tabular}{ccc}
                \hline\hline \rule{0cm}{2.8ex}
                \rule{0cm}{2.8ex}parameter & value & unit\\ 
                \hline
                \rule{0cm}{2.8ex}$P$ & $9.39864^{(a)}$ & $[\mathrm{d}]$\\ 
                \rule{0cm}{2.6ex}$\omega$ & $90$ & $[\mathrm{^\circ}]$\\ 
                \rule{0cm}{2.6ex}$T_p$ & $59007.03 \pm 0.16$ & $[\mathrm{MJD}]$\\ 
                \rule{0cm}{2.6ex}$\gamma$ & $260.2 \pm 15$ & $[\mathrm{km\,s^{-1}}]$\\ 
                \rule{0cm}{2.4ex}$e$ & {$0^{(a)}$} &\\ 
                \rule{0cm}{2.4ex}$K1$ & {$183.9 \pm 11$} &$[\mathrm{km\,s^{-1}}]$\\ 
                \rule{0cm}{2.4ex}$K2$ & {$33.8 \pm 25$} &$[\mathrm{km\,s^{-1}}]$\\ 
                 \hline                 
 
        \end{tabular}
                \rule{0cm}{2.8ex}%
                \tablefoot{
                \ignorespaces 
                $^{(a)}$ Fixed during the analysis
               
                }
        \label{table:rvfit_results}
    \end{table}

\section{Binary evolution calculations with MESA}
\label{app:MESA}

    Our binary evolutionary models presented here are calculated with the MESA code version r23.05.1. Similarly to the stellar evolution model from \citet{Brott2011}, we use tailored abundances for H, He, C, N, O, Mg, Si, and Fe that are in accordance with the baseline abundances of the SMC and the ones used for our spectral analysis.

    In our models, we employ the standard mixing length theory \citep{Boehm1958} using the Ledoux criterion and a mixing length coefficient of $\alpha_\mathrm{mlt}=1.5$ to model convection. Overshooting is included as step overshooting for core-hydrogen burning so that the convective core can extend up to $0.335\,H_\mathrm{P}$ \citep{Brott2011,Schootemeijer2019}. Thermohaline mixing is included with a standard coefficient of $\alpha_\mathrm{th}=1$ \citep{Kippenhahn1980}. Rotational mixing within our models is treated as a diffusive process taking into account Eddington-Sweet circulations, Goldreich-Schibert-Fricke instabilities, as well as dynamical and shear instabilities \citep{Heger2000}. Following \citet{Brott2011}, we set the rotational mixing efficiency coefficients to $f_c=1/30$ and $f_\mu=0.1$. Furthermore, angular momentum transport via magnetic fields is described using the Taylor-Spruit dynamo \citep{Spruit2002}. 
        
    During the OB phase ($X_\mathrm{H}>0.7$) we use the mass-loss recipe of \citet{Vink2001}. Inspired from our comparison of the derived mass-loss rates of partially stripped stars to common mass-loss recipes (see Sect.~\ref{sec:discussion-mdot}), we choose to use the mass-loss recipe of \citet{Vink2017} for stars with $X_\mathrm{H}<0.4$ instead of a WR recipe. For stars in the transition phase, we interpolate linearly between the two mass-loss recipes to avoid sudden jumps in the Hertzsprung-Russell diagram (HRD). Since our stars do not evolve to the RSG phase, no mass-loss recipe is needed for this regime.

    Mass transfer via Roche Lobe overflow (RLOF) in our MESA models is calculated using the inbuilt implicit scheme the so-called ``contact'' scheme. During the mass transfer the accretion on the secondary is treated as rotation-dependent, meaning that as soon as a star reaches its breakup velocity, no further mass can be accreted. Additionally, we assume that mass can be accreted from the stellar wind, however, this fraction is marginal compared to the mass-loss rates and mass-transfer rates during RLOF.

    Our primary stars are calculated until they reach core-carbon depletion. The evolution of the secondary star continues until core-hydrogen depletion, occurring prior to any potential interaction with the companion.
    
\section{Additional plots}
\label{appendix:plots}

\begin{figure*}  

\vspace{0.5cm}
\includegraphics[scale=0.9]{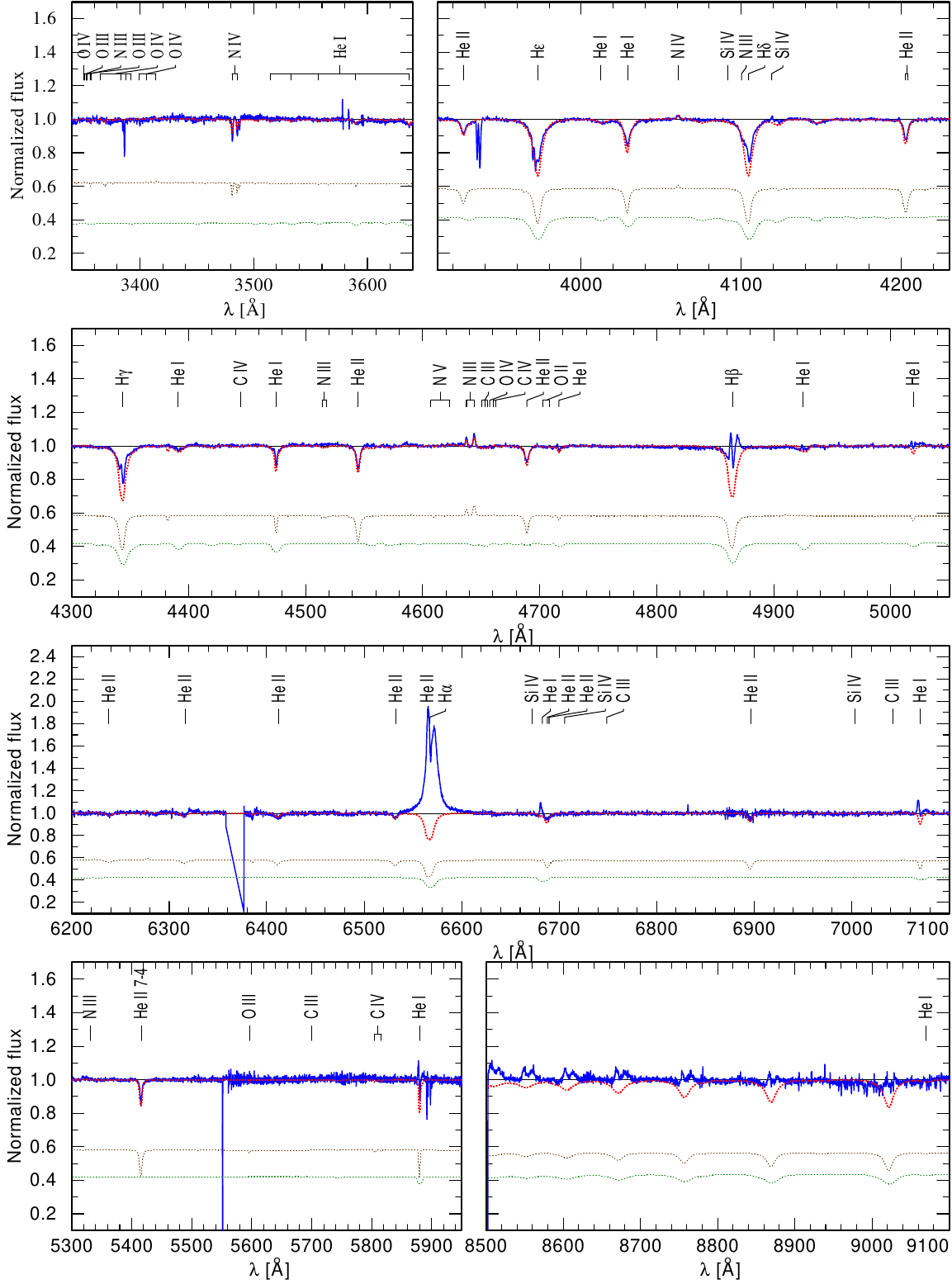}
\caption{Observed spectra of  2dFS\,2553 (solid blue) compared to the model spectra.  The composite
model (dotted red) is the weighted sum of the stripped star primary (dotted brown)
and rapidly rotating Be star secondary (dotted green) model spectra.}
\label{fig:optfit2}
\end{figure*}   

\begin{figure*}  

\vspace{0.5cm}
\includegraphics[scale=0.9,trim={0cm 0cm 0cm 6.5cm},clip]{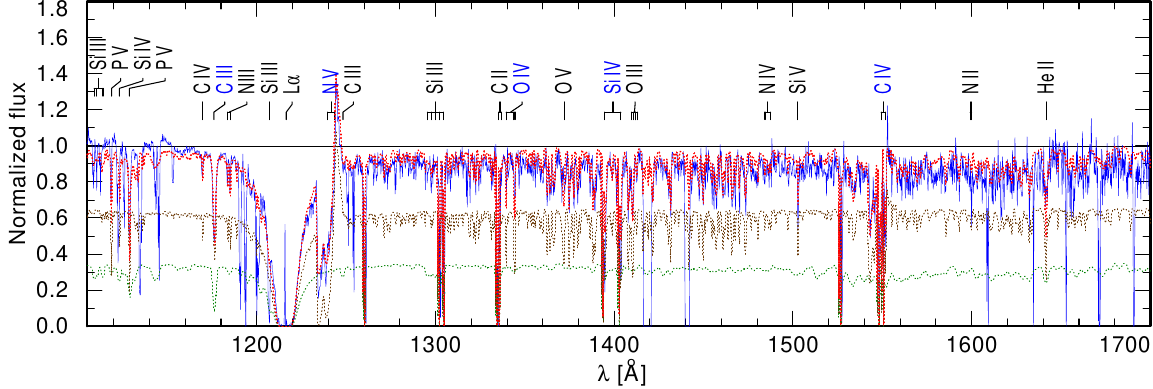}
\caption{Observed HST UV spectra of 2dFS\,2553 (solid blue) compared to the model spectra.  The composite
model (dotted red) is the weighted sum of the stripped star primary (dotted brown)
and rapidly rotating Be star secondary (dotted green) model spectra.}
\label{fig:uvfit2}
\end{figure*}  
 
\begin{figure}[!htb]  

\vspace{0.5cm}
\includegraphics[scale=0.9,trim={0cm 6.7cm 0cm 0cm},clip]{spectraUV-2553.pdf}
\caption{Observed  flux calibrated UV spectra and photometry of 2dFS\,2553 (blue) compared to the model SED. The composite
model (dashed red) is the weighted sum of the stripped star primary (brown) and rapidly rotating B star secondary (green) model.}
\label{fig:sedfit2}
\end{figure}

\begin{figure*}  

\vspace{0.5cm}
\includegraphics[scale=0.9]{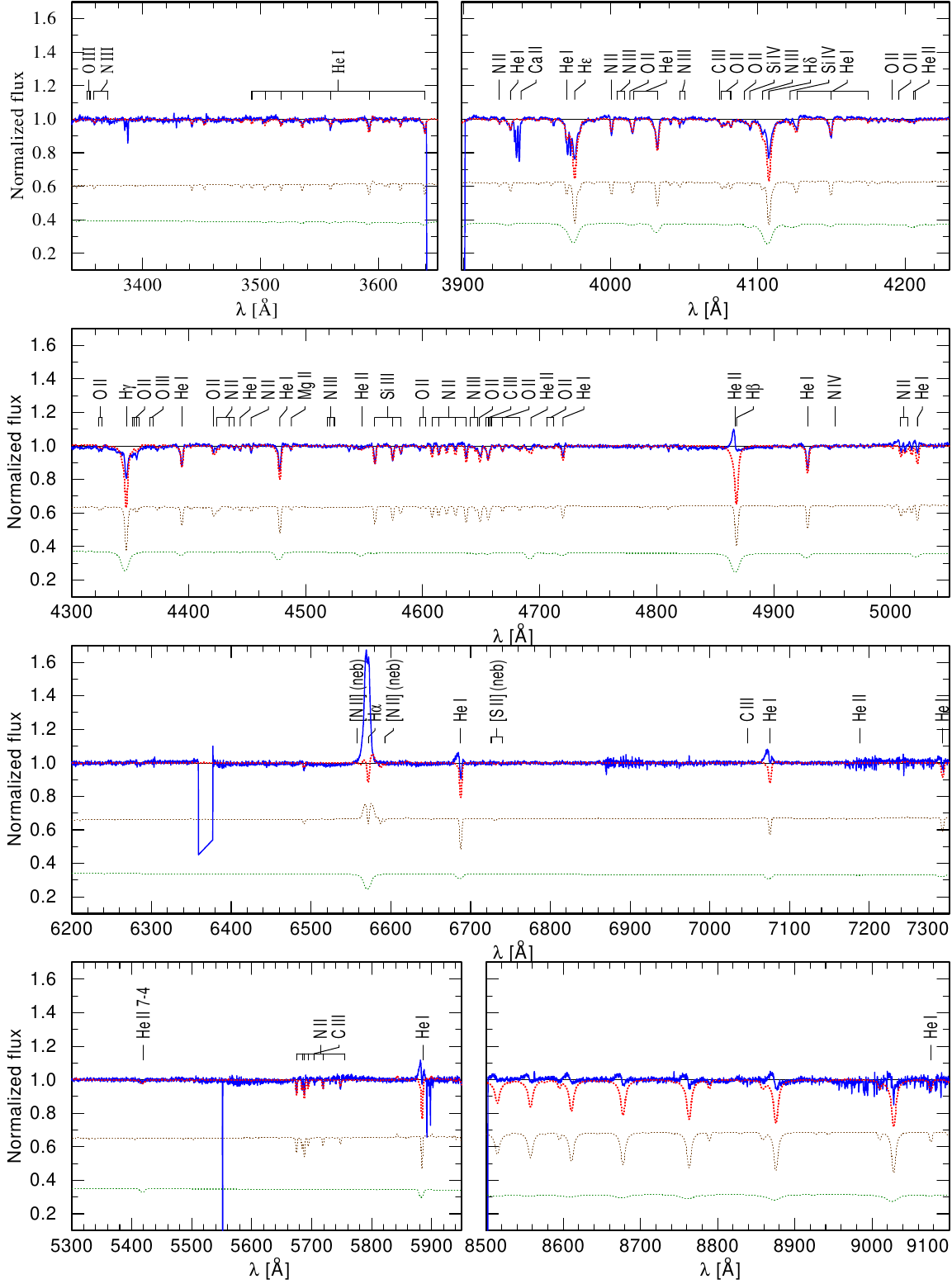}
\caption{Observed spectra of \mbox{Sk\,-71$^{\circ}$\,35} (solid blue) compared to the model spectra.  The composite
model (dotted red) is the weighted sum of the stripped star primary (dotted brown)
and rapidly rotating Oe star secondary (dotted green) model spectra.}
\label{fig:optfit3}
\end{figure*}   

\begin{figure*}  

\vspace{0.5cm}
\includegraphics[scale=0.9,trim={0cm 0cm 0cm 6.5cm},clip]{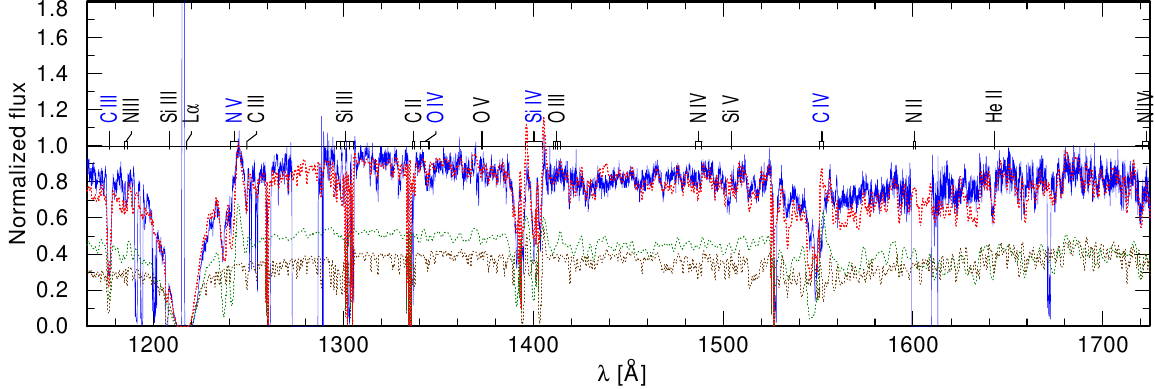}
\caption{Observed HST UV spectra of \mbox{Sk\,-71$^{\circ}$\,35} (solid blue) compared to the model spectra.  The composite
model (dotted red) is the weighted sum of the stripped star primary (dotted brown)
and rapidly rotating Oe star secondary (dotted green) model spectra.}
\label{fig:uvfit3}
\end{figure*}

\begin{figure}[!htb]  

\vspace{0.5cm}
\includegraphics[scale=0.9,trim={0cm 6.7cm 0cm 0cm},clip]{specUV-7135.pdf}
\caption{Observed flux calibrated UV spectra and photometry of  \mbox{Sk\,-71$^{\circ}$\,35} (blue) compared to the model SED. The composite
model (dashed red) is the weighted sum of the stripped star primary (brown) and rapidly rotating B star secondary (green) model.}
\label{fig:sedfit3}
\end{figure}

\end{document}